\documentclass{ar-astro2e}
\usepackage{color}
\usepackage[colorlinks=true,linkcolor=blue,citecolor=blue,urlcolor=blue]{hyperref} 
\usepackage{amsmath,amssymb}
\usepackage{txfonts}
\usepackage{ARAstroBib}
\usepackage{graphicx}
\begin{document}
\jname{Annu. Rev. Astronomy and Astrophysics}
\jyear{2013}
\jvol{1}
\ARinfo{1056-8700/97/0610-00}

\title{Three-Dimensional Dust Radiative Transfer\thanks{This review was the result of
    a collaboration of equals; the ordering of the authorship list is
    not significant.}}

\markboth{Steinacker, Baes, \& Gordon}{3D dust radiative transfer}

\newcommand{\bfx}{{\boldsymbol{x}}}
\newcommand{\bfn}{{\boldsymbol{n}}}
\newcommand{\bfS}{{\boldsymbol{S}}}
\newcommand{\bfj}{{\boldsymbol{j}}}
\newcommand{\txd}{{\text{d}}}
\newcommand{\txe}{{\text{e}}}
\newcommand\mnras{MNRAS}
\newcommand\apj{ApJ}
\newcommand\apjs{ApJS}
\newcommand\apjl{ApJL}
\newcommand\aap{A\&A}
\newcommand\apss{ApSS}
\newcommand\jqsrt{JQSRT}
\newcommand\rmxaa{RMxAA}
\newcommand\araa{ARAA}
\newcommand\na{NewA}
\newcommand\aj{AJ}
\newcommand\icarus{Icarus}
\newcommand\nat{Nature}

\author{J\"urgen Steinacker$^{1,2}$, Maarten Baes$^3$, and Karl D.\ Gordon$^{4,3}$\\
\affiliation{
$^1$UJF-Grenoble 1/CNRS-INSU, Institut de Plan{\'e}tologie et d'Astrophysique de Grenoble \\
(IPAG), UMR 5274, Grenoble, F-38041, France; email: \href{mailto:stein@mpia.de}{stein@mpia.de}\\
$^2$Max-Planck-Institut f\"ur Astronomie, D-69117 Heidelberg, Germany \\
$^3$Sterrenkundig Observatorium, Universiteit Gent, B-9000 Gent, Belgium; \\
email: \href{mailto:maarten.baes@ugent.be}{maarten.baes@ugent.be} \\
$^4$Space Telescope Science Institute, Baltimore, Maryland 21218; \\
email: \href{mailto:kgordon@stsci.edu}{kgordon@stsci.edu}}
}


\begin{abstract}
Cosmic dust is present in many astrophysical objects, and recent observations across the electromagnetic spectrum show that the dust distribution is often strongly three-dimensional (3D). Dust grains are effective in absorbing and scattering ultraviolet (UV)/optical radiation, and they re-emit the absorbed energy at infrared wavelengths. Understanding the intrinsic properties of these objects, including the dust itself, therefore requires 3D dust radiative transfer (RT) calculations. Unfortunately, the 3D dust RT problem is nonlocal and nonlinear, which makes it one of the hardest challenges in computational astrophysics. Nevertheless, significant progress has been made in the past decade, with an increasing number of codes capable of dealing with the complete 3D dust RT problem. We discuss the complexity of this problem, the two most successful solution techniques [ray-tracing (RayT) and Monte Carlo (MC)], and the state of the art in modeling observational data using 3D dust RT codes. We end with an outlook on the bright future of this field.
\end{abstract}

\maketitle

\section{INTRODUCTION}
\label{intro}

Given the dominant role of radiation in astrophysics, its transport through a medium is one of the most fundamental processes to be considered. Analyzing the radiation an object sends provides us with information about not only its radiation source but also the medium in between and surrounding the object and the observer.

Dust grains play a special role in producing and processing radiation. They are efficient at absorbing and scattering UV through near-infrared (NIR) photons and then reradiating the absorbed energy in the infrared and submillimeter (submm) wavelength range. Cosmic dust can be found in many astrophysical objects such as the Solar System \citep{2010ApJ...719.1370H}, comets and meteoroids \citep{2005Natur.437..987K}, substellar atmospheres \citep{2012ApJ...755...67H}, young stellar objects \citep{2008ApJ...684..411K}, protostellar to protoplanetary disks \citep{2009ApJS..180...84W}, evolved stars \citep{2011A&A...526A.162G}, reflection nebulae \citep{2011MNRAS.411.1137C}, supernova remnants \citep{2009ApJ...700..579R}, molecular clouds \citep{2012ApJ...757...59M}, the interstellar medium (ISM) \citep{2008A&A...479..453Z}, galaxies \citep{2011MNRAS.417.1510D}, AGN \citep{2000A&A...354..453H}, and the high-redshift universe \citep{2007ApJ...662..927D}. For an unbiased analysis of these objectsÕ intrinsic quantities, it should be taken into consideration that dust grains modify the radiation fields in these objects. Such an analysis requires performing radiative transfer (RT) calculations.

Aside from its importance as a tracer, the physical and chemical processes related to dust itself are of interest. They cover its formation; its cycle in galaxies; the variation in opacity with chemical composition; its growth and destruction processes in cloud cores and circumstellar disks, which allow it to act as a building block for planets; its interaction with magnetic fields; and its surface chemistry. For example, the dust RT is important for understanding chemistry in the ISM, as photodissociation rates are strongly dependent on the UV radiation field that includes a significant amount of photons scattered from dust grains. This review discusses the physical properties of dust only in the context of RT and the modeling of objects; for other aspects mentioned above, we refer the reader to any one of the many published works on dust \citep[e.g.][]{2003ARA&A..41..241D,2009ASPC..414.....H,2010ARA&A..48...21H}.

\begin{figure}
\centering
\includegraphics[width=\textwidth]{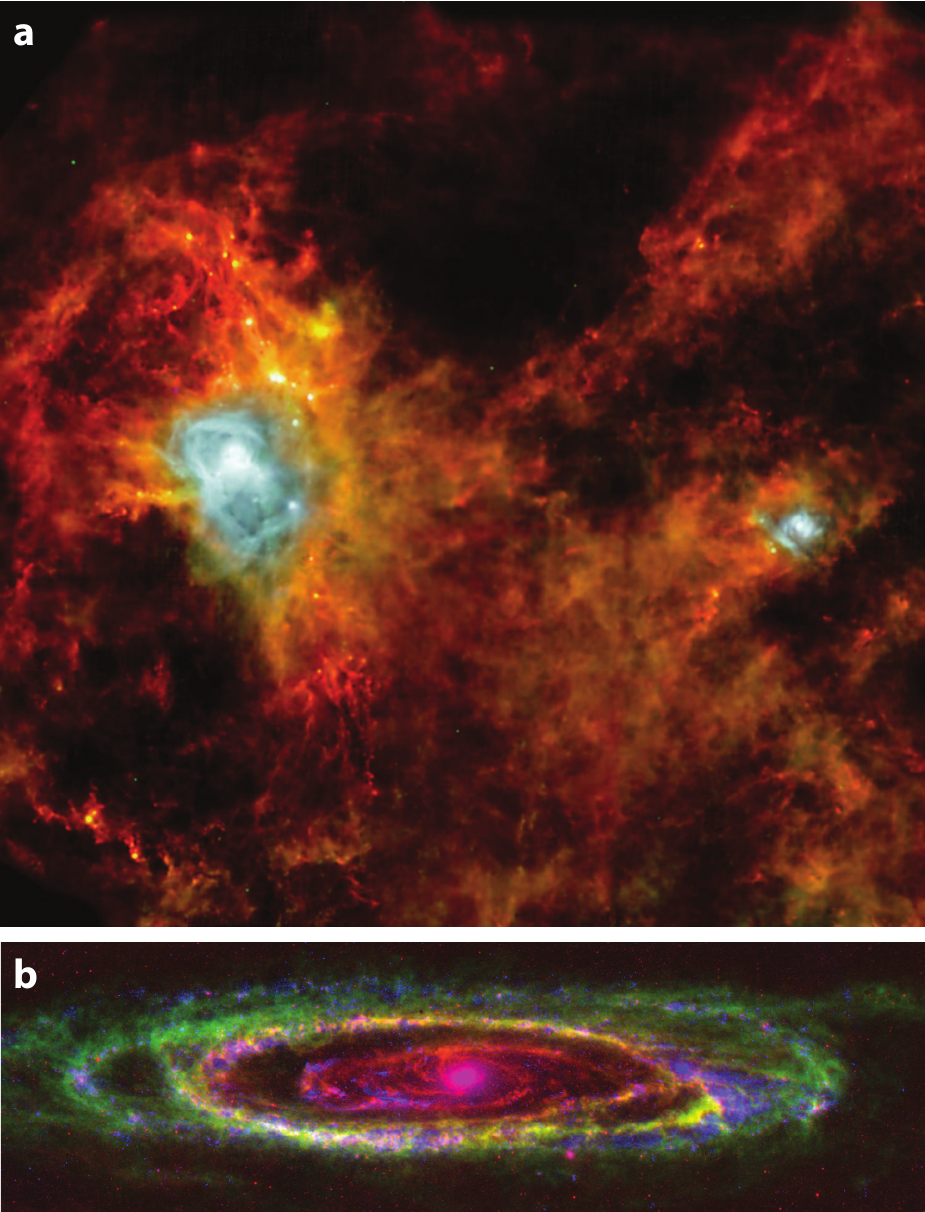}
\caption{The complex and filamentary structures of ISM dust are seen clearly in both Milky Way star-formation regions ({\em{a}}) and external galaxies ({\em{b}}). Panel {\em{a}} is a color image of the Aquila star-formation complex \citep{2010A&A...518L.102A}. This image was taken as part of the Gould Belt Survey Herschel Key Project; covers $\sim$11 deg$^2$; and was created from PACS 70-$\mu$m ({\em{blue}}), PACS 160-$\mu$m ({\em{green}}), and SPIRE 250-$\mu$m ({\em{red}}) observations. Panel {\em{b}} shows the complex structure of the ISM in M31 from H{\sc{i}} observations ({\em{green}}; see also \citealt{2009ApJ...695..937B}), embedded star formation using {\em{Spitzer}} 24-$\mu$m data ({\em{red}}; see also \citealt{2006ApJ...638L..87G}), and unobscured young stars from the GALEX far-UV images ({\em{blue}}; see also \citealt{2005ApJ...619L..67T}). Abbreviations: GALEX, the {\em{Galaxy Evolution Explorer}}; ISM, interstellar medium; PACS, Photodetector Array Camera \& Spectrometer; SPIRE, Spectral and Photometric Imaging Receiver; {\em{Spitzer}}, the {\em{Spitzer Space Telescope}}.}
\label{fig_aquila_structure}
\end{figure}

Many dusty objects have been observed at increasingly higher spatial resolution in the past 10 years. These observations cover UV/optical/NIR [e.g., the {\em{Hubble Space Telescope}} (HST), the {\em{Galaxy Evolution Explorer}} (GALEX), and ground-based telescopes] and infrared/submillimeter [e.g., the {\em{Spitzer Space Telescope}} (Spitzer), {\em{Akari}}, {\em{Herschel}}, {\em{Planck}}, the {\em{Wide-Field Infrared Survey Explorer}} (WISE), the {\em{Atacama Large Millimeter Array}} (ALMA), the {\em{James Clerk Maxwell Telescope}} (JCMT), the {\em{Atacama Pathfinder Experiment}} (APEX), and the {\em{Institut de Radio Astronomie Millim\'etrique}} (IRAM) 3-m Telescope] wavelengths. Space telescopes exploring atmospherically absorbed wavelength windows, high-resolution interferometric data, polarization data, and all-sky maps are just a few examples of the rich data set that awaits the RT modeler. Comparison of dust RT calculations with global and pixel-by-pixel resolved spectral energy distributions (SEDs) of dusty objects provides information on the properties of the illuminating sources (stars, accretion disks, integrated star formation rate, etc.), the distribution of the dust (disk structures, cloud geometries, underlying multiphase nature, etc.), and properties of the dust grains (size, shape, and composition). A feature commonly seen in high spatial resolution images at all wavelengths is the complex nature of the dust density distribution. Examples of global 3D geometries include complex arm structures in spiral galaxies \citep{2006A&A...458..441P, 2012A&A...546A..34F}, large scale filaments in star-forming regions \citep{2010A&A...518L.102A, 2011A&A...529L...6A}, and bow-shocked shells around evolved stars \citep{2012A&A...537A..35C}. A prominent example of locally complex 3D geometries is the known fractal nature of the ISM \citep{1987Ap&SS.133..193B, 1991ApJ...378..186F}. Images illustrating complex local and global 3D dust structures are shown in {\bf{Figure~{\ref{fig_aquila_structure}}}}. In addition, the illuminating sources of dust have been long known to have complex distributions from the combination of the anisotropic interstellar radiation field and local neighboring stars to the stellar distribution in galaxies. Both of these issues show that a complete 3D treatment of RT is inevitable and critical for progress in many fields.

Among the many computational problems in astrophysics, 3D line and dust RT has long been a major challenge and often approximated or neglected. Although, for example, 3D magnetohydrodynamics (MHD) codes have existed for many years, radiation transport is considered to be one of the four grand challenges in computational astrophysics. [See, for example, the {\em{Grand Challenge Problems in Computational Astrophysics}} conference series, 2005--2007, at the Institute for Pure and Applied Mathematics at the University of California, Los Angeles ({\url{https://www.ipam.ucla.edu/}}).]. Dust RT is different than line RT in that the dust opacities generally do not depend on the RT solution itself.

The reasons to neglect or approximate 3D dust RT are manifold. A good portion of the difficulties arises because the underlying physical processes combine, in the stationary case, to a nonlocal and nonlinear 6D problem. Because the radiation field needs to be determined in all directions, at any spatial location, and for each wavelength, the solution vector itself comprises three dimensions more than the variables in MHD problems. The RT problem is nonlocal in space (propagation of the photons within the entire domain), direction (scattering and absorption/re-emission), and wavelength (absorption/re-emission). This nonlocality makes it difficult to simplify the problem by neglecting processes or wavelengths. For example, absorption and scattering have roughly the same efficiency from UV to NIR, with strongly anisotropic scattering \citep{Gordon04Review}. In modeling far-infrared images, a consistent treatment of the dust emission requires the RT to be calculated where the dust absorption happens, at shorter wavelengths. Therefore, most of the current 3D dust applications are intrinsically multiwavelength in nature.

Other difficulties are related to the complexity of 3D structures. The underlying grids to resolve the sink and source contributions to the radiation fields are generally discretized and require substantial storage, and the RT calculation effort rises with the decreasing cell size. Moreover, when modeling the structure, the spatial distribution of the sources and sinks has to be parametrized. Modeling complex structures with a simple spatial distribution model can lead to misleading results. \citet{1996ApJ...463..681W} showed that RT through a 3D fractal dust distribution was significantly different than through similar, yet smooth, distributions. In addition to the longer runtimes expected for the 3D dust RT code, an exploration or optimization of the parameter space challenges the capabilities of current computers. Finally, more than for simple geometries, applying 3D dust RT is challenging given the loss of information due to projection effects.

As a result of nonlocal and nonlinear effects, the radiative transfer equation (RTE) is an integro-differential equation including a scattering integral; the thermal source term is nonlinearly coupled to a double-integral equation, making it difficult to apply common solvers. Moreover, the spatially varying extinction causes changes in the numerical nature of the RTE. Its character changes from parabolic for the diffusive transport to hyperbolic for freely streaming photons, to a combination of the two in the numerically difficult transition region. Solving such a high-dimensional nonlocal, nonlinear problem requires substantial computational resources (both computing power and memory), affecting the solution algorithms and potentially limiting the model complexity.

We review the significant progress made in this evolving and dynamic field to tackle this grand challenge problem. Within the past 10 years, the availability of high resolution images and increase in computer speed and storage have triggered an expansion of the dust RT community and the development of new codes capable of dealing with the complete 3D dust RT problem. Many of the techniques used to solve the 3D dust RT problem were developed originally for 1D or 2D geometries. The added computational complexity of solving the 3D problem has emphasized the need for highly efficient techniques, leading to refinements in and use of all possible 1D/2D methods in most 3D codes. Applications explicitly using 3D RT include models of young stellar objects \citep{1998A&A...340..103W}, protostellar to protoplanetary disks \citep{2006ApJ...636..362I, 2006AA...456....1N}, reflection nebulae \citep{1996ApJ...463..681W}, molecular clouds \citep{2005A&A...434..167S, 2009AA...502..833P}, spiral galaxies \citep{2008AA...490..461B, 2012ApJ...746...70S}, interacting and starburst galaxies \citep{2007ApJ...658..840C, 2011ApJ...743..159H}, and AGN \citep{2008A&A...482...67S, 2012MNRAS.420.2756S}. This expansion motivated the ÒCosmic Dust and Radiative TransferÓ workshop held in Heidelberg in 2008 (\url{http://www.mpia.de/RT08/}). During the workshop, we realized that a review of the various techniques and applications addressing 3D dust RT was needed to communicate common strategies between related fields. It would also be useful for coders and users of dust RT codes and people wishing to enter the field (including writers of line RT and MHD codes).

Although various solver techniques are used for RT problems, 3D dust RT is commonly solved using the Monte Carlo (MC) technique, with some applications using the ray-tracing (RayT) technique. Because modern MC solvers make use of some RayT methods, this review focuses on the spectrum of techniques based on these two approaches.

Other RT solution methods exist but have either not been used to solve the complete 3D dust RT problem, or have shown clear disadvantages. One potential solution method is discretizing the RTE, for example, with a finite-difference approach in spatial Cartesian coordinates and in direction space, to create a system of linear equations \citep{1991JQSRT..45...47S,2002JQSRT..75..765S}. The corresponding matrix is extremely difficult to solve even with powerful matrix solvers \citep{1992SIAMJSSC..13...631V}. Two more methods are used for RT problems, but have yet to be applied to 3D dust RT. The discretization can be performed on unstructured grids (e.g., a Delaunay grid), and this can provide a very fast solver. Currently, such algorithms have been updated to handle freely streaming photon packages and changes in the optical depth, but treatment of scattering has yet to be explored. Finally, the moment method expands the intensity as a function of angle using spherical harmonics as basis functions. It has several numerical advantages both in terms of solution accuracy and storage requirements, but can exhibit nonphysical oscillations. A common variant of the moment method is the variable Eddington tensor method (used, e.g., for 2D in the code RADICAL; see description in \citealt{2004AA...417..793P}).

This review starts with the mathematical definition of the full 3D dust RT problem. Next, we present the discretization of the problem in spatial, direction, and wavelength dimensions. The RayT and MC methods of solving the 3D RT problem are described in detail. Challenges in comparing RT models with observations are discussed. A listing of existing 3D dust RT codes is given along with current benchmarking efforts. Finally, the review is concluded with a summary and discussion of the future of 3D RT.

\section{THE THREE-DIMENSIONAL DUST RADIATIVE TRANSFER PROBLEM}\label{radia}

\subsection{The Radiative Transfer Equation}

The stationary radiation field is described by the specific intensity $I(\bfx,\bfn,\lambda)$, where $\bfx$ is the location in space, $\bfn$ is a unit vector indicating the direction of the radiation, and $\lambda$ is its wavelength. The specific intensity represents the amount of energy carried by radiation in a unit wavelength interval, which is transported per unit solid angle and per unit time across an element of unit area perpendicular to $\bfn$. (The specific intensity can be defined as the intensity per unit of wavelength or per unit of frequency. The typical convention is to indicate it with a subscript, i.e., as $I_\lambda$ or $I_\nu$. We adopt the per unit of wavelength convention and drop the subscript so as not to overload the notations.) The continuum RTE describes how the specific intensity varies as a result of interactions with a medium filled with sources and sinks. In its general form, it can be written as  \citep[see
e.g.,][]{1960ratr.book.....C, 1979rpa..book.....R}
\begin{equation}
  \bfn\cdot\nabla 
  I(\bfx,\bfn,\lambda)
  =
  -
  \kappa(\bfx,\lambda)\,\rho(\bfx)\,
  I(\bfx,\bfn,\lambda)
  +
  j(\bfx,\bfn,\lambda).
  \label{RTE-General}
\end{equation}
The left-hand side of this equation represents the change in intensity over an infinitesimal distance along the path determined by the position $\bfx$ and the propagation direction $\bfn$. The first term on the right-hand side represents the extinction, i.e., the loss of radiant energy, when radiation passes through matter. Here, $\kappa(\bfx,\lambda)$ is the mass extinction coefficient, and $\rho(\bfx)$ is the mass density. The second term on the right-hand side represents the source term, i.e., the new luminosity released into the medium at $\bfx$ in direction $\bfn$. The complexity of the RTE depends on the nature of the source and sink terms, i.e., the different physical processes that are responsible for extinction and emission.

An alternative form of the RTE (Equation \ref{RTE-General}) uses explicitly the distance $s$ along the path defined by a position $\bfx$ and propagation direction $\bfn$ as a variable. We then obtain
\begin{equation}
  \frac{\txd I}{\txd s}(s,\lambda)
  =
  -\kappa(s,\lambda)\,\rho(s)\,I(s,\lambda)
  +
  j(s,\lambda).
  \label{RTE-GeneralPath}
\end{equation} 
If we assume for now that the source term $j$ does not depend on the intensity $I$, we can readily solve the differential equation
\begin{equation}
  I(s,\lambda)
  =
  \int_{-\infty}^s j(s',\lambda)\, \txe^{-\tau(s',s,\lambda)}\,
  \txd s'\, ,
  \label{RTE-FormalSolution}
\end{equation}
with the optical depth between two positions defined as
\begin{equation}
  \tau(s_1,s_2,\lambda)
  =
  \int_{s_1}^{s_2}
  \kappa(s,\lambda)\,\rho(s)\,\txd s.
\end{equation}
Equation~\ref{RTE-FormalSolution} has a simple physical interpretation: It shows that the intensity at any position $s$ along a path results from the emission at all anterior points $s'$ along the path, reduced by a factor of $\txe^{?\tau(s,s',\lambda)}$ to account for the extinction from the intervening matter. We stress that Equation~\ref{RTE-FormalSolution} is only a formal solution of the RTE but not a very useful one. Indeed, the emissivity generally does not only depend on position, direction, and wavelength, but also on the specific intensity itself. The formal solution (Equation~\ref{RTE-FormalSolution}) is then no more than an integral equation of the RTE itself; this is particularly the case for dust RT. The remainder of Section~{\ref{radia}} gradually expands the dust RTE by including the various relevant physical processes.

\subsection{Primary Emission and Absorption}

Primary emission and absorption in a dusty medium are two important and obvious processes to take into account. Primary emission accounts for the radiative energy added to the radiation field-- this is often stellar emission, but it can also include, for example, radiation from an AGN, emission line radiation from ionized gas, or Bremsstrahlung. In general form, it can be characterized by a function $j_*(\bfx,\bfn,\lambda)$. Absorption is the process in which electromagnetic radiation is taken up by dust grains and transformed into the internal energy. It is characterized by the absorption coefficient $\kappa_{\text{abs}}$; for a given chemical composition, size, and shape of a dust grain, the absorption coefficient can in principle be determined at any wavelength \citep{ANDP:ANDP19083300302, 1973ApJ...186..705P, 1988ApJ...333..848D, 2002sael.book.....M, 2005A&A...432..909M}.

When we take only primary emission and absorption by dust into account, the RTE (Equation~\ref{RTE-General}) becomes
\begin{equation}
  \frac{\txd I}{\txd s}(\bfx,\bfn,\lambda)
  =
  -\kappa_{\text{abs}}(\bfx,\lambda)\,\rho(\bfx)\,I(\bfx,\bfn,\lambda)
  +
  j_*(\bfx,\bfn,\lambda).
  \label{RTE-EmissionAbsorption}
\end{equation}
This equation is a simple first-order differential equation, which can be solved by simply integrating along the line of sight, as is done for the formal solution (Equation~\ref{RTE-FormalSolution}). For general 3D geometries, this integration is done numerically.

\subsection{Including Scattering}
\label{sec_RT_dust_emission}
 
The complexity of the RTE increases substantially when we take scattering into account. Scattering, as with absorption, removes radiation from a beam and hence accounts for a second sink term in the RTE, the efficiency of which is quantified by the scattering coefficient $\kappa_{\text{sca}}$. Rather than converting the radiation to internal energy, it emits the same radiation in a different direction. Scattering therefore does not only imply a second sink term, but also a second source term. The scattering phase function $\Phi(\bfn, \bfn', \bfx, \lambda)$ describes the probability that a photon originally propagating in direction $\bfn'$ and scattered at position $\bfx$, will have $\bfn$ as its new propagation direction after the scattering event. Given this definition, the phase function satisfies the normalization
\begin{equation}
  \int_{4\pi} \Phi(\bfn, \bfn', \bfx, \lambda)\, \txd\Omega'
  =
  \int_{4\pi} \Phi(\bfn, \bfn', \bfx, \lambda)\, \txd\Omega
  =
  1.
\end{equation}
Adding to the RTE the sink and source terms due to scattering, we obtain
\begin{multline}
  \frac{\txd I}{\txd s}(\bfx,\bfn,\lambda)
  =
  -
  \kappa_{\text{ext}}(\bfx,\lambda)\,\rho(\bfx)\,
  I(\bfx,\bfn,\lambda)
  +
  j_*(\bfx,\bfn,\lambda)
  \\
  +
  \kappa_{\text{sca}}\,(\bfx,\lambda)\,\rho(\bfx)
  \int_{4\pi}
  \Phi(\bfn,\bfn',\bfx,\lambda)\,
  I(\bfx,\bfn',\lambda)\,
  \txd\Omega'\, ,
  \label{RTE-Scattering}
\end{multline} 
where the extinction coefficient $\kappa_{\text{ext}} = \kappa_{\text{abs}} + \kappa_{\text{sca}}$. Unlike the simple differential equation (Equation~\ref{RTE-EmissionAbsorption}), Equation~\ref{RTE-Scattering} is an integro-differential equation in which the radiation fields at all different positions and in all different directions are coupled.

In many RT calculations, scattering is important and adds significant complexity, given that dust scattering is anisotropic. This is especially true for UV to NIR wavelengths: Observations indicate that the dust scattering albedo at these wavelengths is at least 50\% and that scattering off dust grains is strongly anisotropic \citep{1970A&A.....9...53M, 1995ApJ...446L..97C, 2002ApJ...575..240B, Gordon04Review}. [\url{http://www.stsci.edu/~kgordon/Dust/Scat_Param/scat_data.html} contains the up-to-date versions of the original plots of albedo and scattering phase function asymmetry versus wavelength from \citet{Gordon04Review}.] Using an isotropic phase function or other approximations such as an isotropic two-stream approximation or a (effective) forward scattering \citep[see e.g.,][]{1973IAUS...52..505C, 1984ApJ...287..228N, 1994ApJ...429..582C} might not always be physically justifiable. Several authors demonstrated that improper treatment of anisotropic scattering leads to significant errors \citep[e.g.][]{1988ApJ...333..673B, 1992ApJ...393..611W, 2001MNRAS.326..733B}. Even for radiation at mid-infrared (MIR) wavelengths or longer, where the scattering off common ISM grains is low, an application calculating the heating of the grains will need to properly handle the scattering \citep{2012A&A...547A..11N}.

The Henyey-Greenstein (HG) phase function \citep{1941ApJ....93...70H} is the most widely used parametrization of the dust phase function and provides a good single-parameter approximation. Dust grain models predict small deviations from an HG phase function, and other parametrizations or numerical phase functions can be used for increased accuracy \citep{1975JQSRT..15..839K, 1985AA...146...67H, 2003ApJ...598.1017D}.

\subsection{Radiative Transfer in Dust Mixtures}

In any real dust medium, there is a range of different types of dust grains present, with various chemical compositions, sizes, shapes, and number densities. Each grain type $i$ is characterized by its own absorption coefficient $\kappa_{\text{abs},i}(\lambda)$, scattering coefficient $\kappa_{\text{sca},i}(\lambda)$, and scattering phase function $\Phi_i(\bfn, \bfn',\lambda)$. If we denote the relative contribution of each grain type $i$ at the position $\bfx$ to the total dust number density as $w_i(\bfx)$, the RTE becomes
\begin{multline}
  \frac{\txd I}{\txd s}(\bfx,\bfn,\lambda)
  =
  -
  \sum_i w_i(\bfx)\,
  \kappa_{\text{ext},i}(\lambda)\,\rho(\bfx)\,
  I(\bfx,\bfn,\lambda)
  +
  j_*(\bfx,\bfn,\lambda)
  \\
  +
  \sum_i w_i(\bfx)\,
  \kappa_{\text{sca},i}(\lambda)\,\rho(\bfx)\,
  \int_{4\pi}
  \Phi_i(\bfn,\bfn',\lambda)\,
  I(\bfx,\bfn',\lambda)\,
  \txd\Omega'\, .
  \label{RTE-Mixtures}
\end{multline} 
Clearly, Equation~\ref{RTE-Mixtures} is formally identical to Equation~\ref{RTE-Scattering}, if we define the absorption coefficient, scattering coefficient, extinction coefficient, and scattering phase function of a dust mixture as, respectively,
\begin{subequations}
\label{meangrain}
\begin{gather}
\kappa_{\text{abs}}(\bfx,\lambda) = \sum_i
w_i(\bfx)\,\kappa_{\text{abs},i}(\lambda),
    	\\
    	\kappa_{\text{sca}}(\bfx,\lambda) = \sum_i
   	w_i(\bfx)\,\kappa_{\text{sca},i}(\lambda),
    \\
    \kappa_{\text{ext}}(\bfx,\lambda) = \sum_i
    w_i(\bfx)\,\kappa_{\text{ext},i}(\lambda),
\end{gather}
and
\begin{equation}  
    \Phi(\bfn, \bfn', \bfx, \lambda) = \frac{\sum_i
      w_i(\bfx)\,\kappa_{\text{sca},i}(\lambda)\, \Phi_i(\bfn,
      \bfn',\lambda)} {\sum_i
      w_i(\bfx)\,\kappa_{\text{sca},i}(\lambda)}.
\end{equation}
\end{subequations}
In terms of primary emission, absorption, and scattering, RT in dust mixtures is completely identical to RT in a dust medium with a single average dust grain, and no approximations are needed \citep{1978cdii.book.....M, 2003ApJ...582..859W}.

\subsection{Including Dust Emission}

In addition to primary emission, absorption, and scattering, a fourth physical process in dust RT, the thermal emission by the dust itself, must be taken into account. Dust grains that absorb radiation re-emit the acquired radiative energy at wavelengths longward of $\sim$1 $\mu$m. To account for this astrophysical process, we need to incorporate the third source term $j_{\text{d}}(\bfx,\lambda)$ in our RTE:
\begin{multline}
  \frac{\txd I}{\txd s}(\bfx,\bfn,\lambda)
  =
  -
  \kappa_{\text{ext}}(\bfx,\lambda)\,\rho(\bfx)\,
  I(\bfx,\bfn,\lambda)
  +
  j_*(\bfx,\bfn,\lambda)
  +
  j_\txd(\bfx,\lambda)
  \\
  +
  \kappa_{\text{sca}}\,(\bfx,\lambda)\,\rho(\bfx)
  \int_{4\pi}
  \Phi(\bfn,\bfn',\bfx,\lambda)\,
  I(\bfx,\bfn',\lambda)\,
  \txd\Omega'\, .
  \label{RTE-ThermalEmission}
\end{multline} 
The dust emissivity term could be perceived as merely an extra source term similar to the primary stellar emissivity term. Its exact form is strongly dependent on which physical emission processes are important, and often it depends, in a complicated and nonlinear way, on the intensity of the radiation field itself.

A common assumption is that the dust grains are in thermal equilibrium with the local interstellar radiation field. In this case, the emissivity of the population of grains of type $i$ can be written as a modified blackbody emission characterized by the equilibrium temperature $T_i(\bfx)$. Summing over all populations, we obtain
\begin{equation}
  j_\txd(\bfx,\lambda)
  =
  \sum_i
  w_i(\bfx)\,
  \kappa_{\text{abs},i}(\lambda)\,\rho(\bfx)\,
  B[T_i(\bfx),\lambda],
  \label{dustemissivity}
\end{equation}
where $B(T,\lambda)$ is the Planck function. The equilibrium temperature of each type of grain is determined by the balance equation, i.e., the condition that the total amount of energy absorbed equals the total amount of emitted energy:
\begin{equation}
\label{balance}
  \int_0^\infty
  \kappa_{\text{abs},i}(\lambda)\,
  J(\bfx,\lambda)\,
  \txd\lambda
  =
  \int_0^\infty
  \kappa_{\text{abs},i}(\lambda)\,
  B[T_i(\bfx),\lambda]\,
  \txd\lambda,
\end{equation}
where $J(\bfx,\lambda)$ represents the mean intensity of the radiation
field,
\begin{equation}
\label{meanintensity}
  J(\bfx,\lambda)
  =
  \frac{1}{4\pi}
  \int_{4\pi} I(\bfx,\bfn,\lambda)\,\txd\Omega.
\end{equation}
The equilibrium temperature of the dust grains depends explicitly on their size and chemical composition. At the same location, dust grains of different sizes and/or chemical compositions will obtain different equilibrium temperatures. So far, we have easily combined the absorption and scattering due to different kinds/sizes of dust grains in the RTE without any approximations. The same cannot be done for the thermal re-emission term. One could compute the average temperature for the different grains on the basis of Equation~{\ref{meangrain}} and obtain a mean temperature. This results in reducing the complexity of the dust mixture at a given position to a single mean grain that will reach a single equilibrium temperature. Although this could be useful, sufficient, or even necessary for certain applications, this is a physically incorrect simplification of the RT problem \citep[e.g.][]{2003ApJ...582..859W}.

Although the assumption of thermal equilibrium is useful in some applications, it breaks down in others. Particularly important is the case where the dust medium contains very small dust grains [including polycyclic aromatic hydrocarbons (PAHs)]. Large dust grains reach thermal equilibrium and emit as modified blackbodies with an equilibrium temperature. However, small dust grains have small heat capacities, and the absorption of even a single UV/optical photon can substantially heat the grain. These small grains will not reach the equilibrium temperature but will instead undergo temperature fluctuations that lead to grain emission at temperatures well in excess of the equilibrium temperature. The emission from small grains is necessary to explain the observed MIR emission of many objects \citep[see, e.g.,][]{1984ApJ...277..623S, 1988ApJ...330..964B, 2000ApJ...532L..21H, 2007ApJ...656..770S, 2007ApJ...663..866D}. When including such transiently heated dust grains, the dust emissivity changes from Equation~\ref{dustemissivity} to
\begin{equation}
  j_\txd(\bfx,\lambda)
  =
  \sum_i
  w_i(\bfx)\,
  \kappa_{\text{abs},i}(\lambda)\,\rho(\bfx)
  \left[
    \int_0^\infty
    P_i(T,\bfx)\,B(T,\lambda)\,
    \txd T
  \right].
\end{equation}
Here, $P_i(T,\bfx)$ is the temperature distribution for dust grains of type $i$ at the location $\bfx$. This temperature distribution depends on the chemical composition and size of the dust grains, as well as on the intensity and hardness of the radiation field in which it is embedded. Several methods have been developed to calculate the temperature distribution of small dust grains, using either matrix operations or time averages \citep[see, e.g.,][]{1986ApJ...302..363D, 1986AA...160..295D, 1989ApJ...345..230G, 1992AA...266..501S, 2001ApJ...551..807D, 2011AA...525A.103C}. The result is that the dust source term is an intricate, nonlinear function of the specific intensity, which adds significant complexity to the RT problem.

Finally, thermal emission is not the only emission process of dust grains: Additional nonthermal processes are extended red emission \citep{1990ApJ...355..182W, 2002ApJ...565..304S} and blue luminescence \citep{2004ApJ...606L..65V}. These nonthermal processes can account for a substantial fraction of the surface brightness of interstellar clouds at optical wavelengths \citep[see, e.g.,][]{1998ApJ...498..522G, 2008ApJ...679..497W}. Both processes can be included as an additional term in the dust source term $j_{\text{d}}(\bfx,\lambda)$ in Equation~\ref{RTE-ThermalEmission}.

\subsection{Radiative Transfer of Polarized Radiation}

The specific intensity $I(\bfx,\bfn,\lambda)$ is not a complete description of the radiation field, as it describes only unpolarized light. Scattering of photons from dust grains naturally produces polarized radiation \citep[see, e.g.,][figure 2]{2001ApJ...551..269G}. In addition, aligned dust grains also produce polarized radiation \citep{2002ApJ...574..205W}. Although alignment of grains has been demonstrated observationally for many decades, the physical mechanism for grain alignment is still a matter of debate \citep{2007JQSRT.106..225L}.

The most common description of polarized radiation uses the Stokes vector $\bfS = (I, Q, U, V )$, where $I$ is the total specific intensity, $Q$ and $U$ the linearly polarized intensity in two axes rotated 45$^\circ$ from one another, and $V$ the circularly polarized intensity. The four components do not form a preferred basis of this space, but were chosen because they can be easily measured or calculated.

The RT formalism we describe above can be extended to include polarized radiation. Instead of a single RTE, we then obtain a vector RTE, or equivalently, a set of four coupled scalar equations:
\begin{multline}
  \frac{\txd\bfS}{\txd s}
  (\bfx,\bfn,\lambda)
  =
  -
  \kappa_{\text{ext}}(\bfx,\lambda)\,
  \rho(\bfx)\,
  \bfS(\bfx,\bfn,\lambda)
  +
  \bfj_*(\bfx,\bfn,\lambda)
  +
  \bfj_\txd(\bfx,\lambda)
  \\
  +
  \kappa_{\text{sca}}(\bfx,\lambda)\,
  \rho(\bfx)\,
  \int_{4\pi}
  {\cal{M}}(\bfn,\bfn',\bfx,\lambda)\,
  \bfS(\bfx,\bfn',\lambda)\,
  \txd\Omega'\, .
  \label{RTEpol}
\end{multline} 
The first complication is the scattering source term, where a $4\times4$ scattering (or Mueller) matrix ${\cal{M}}(\bfn,\bfn',\bfx,\lambda)$, which describes the changes in the Stokes vector when radiation is scattered from propagation direction $\bfn'$ to a new propagation direction $\bfn$, replaces the phase function $\Phi(\bfn,\bfn',\bfx,\lambda)$. For a full description, see e.g., \citet{1983asls.book.....B}; \citet{1994AA...284..187F}; or \citet{1995ApJ...441..400C}. When we consider RT of polarized light through a dust mixture, each type $i$ of grain is characterized by its own Mueller matrix ${\cal{M}}_i(\bfn,\bfn' ,\lambda)$. The RTE can then still be written as Equation~{\ref{RTEpol}}:
\begin{equation}
  {\cal{M}}(\bfn,\bfn',\bfx,\lambda)
  =
  \frac{\sum_i
    w_i(\bfx)\,
    \kappa_{\text{sca},i}(\lambda)\,{\cal{M}}_i(\bfn,\bfn',\lambda)}
  {\sum_i w_i(\bfx)\,\kappa_{\text{sca},i}(\lambda)}.
\end{equation}
A second complication is the thermal emission term: The full Stokes vector is used for thermal emission from aligned grains.

\section{THE DISCRETE THREE-DIMENSIONAL DUST RADIATIVE TRANSFER PROBLEM}
\label{setup}

A general analytical solution of the stationary 3D dust RTE is not possible for any of the non-symmetric applications mentioned in the introduction. To apply numerical solution techniques, solvers generally discretize the solution vector or the physical properties in the RTE. The quantities requiring discretization are the three spatial coordinates, the two directional coordinates, the wavelengths, and/or the dust properties.

A major concern when solving a 6D integro-differential equation is the size of the solution vector. With a resolution of 100 points in each variable, the intensity vector has $10^{12}$ entries. Handling this intensity vector requires an enormous amount of computer memory and speed. Currently, many solution algorithms avoid this requirement by not storing the full solution vector. There are applications for which the full direction-dependent intensity is needed (e.g., when the radiation pressure impacts the gas kinematics). Due to this high dimensionality of 3D RT, the choice of appropriate grids is mandatory to effectively apply existing solution techniques and minimize memory usage and runtime. The solution techniques used do influence the choice of grids (e.g., RayT versus MC).

Another concern is that for a discrete solution vector, the physics is only solved at the grid resolution, even if the solver used has no intrinsic error. Thus, if the grid is too coarse, a strong change in the intensity due to a change from optically thick to thin media may not be resolved, and the derived intensity would have large systematic errors. This makes the choice of grid crucial.

\subsection{Spatial Grids}

\begin{figure}
\centering
\includegraphics[width=\textwidth]{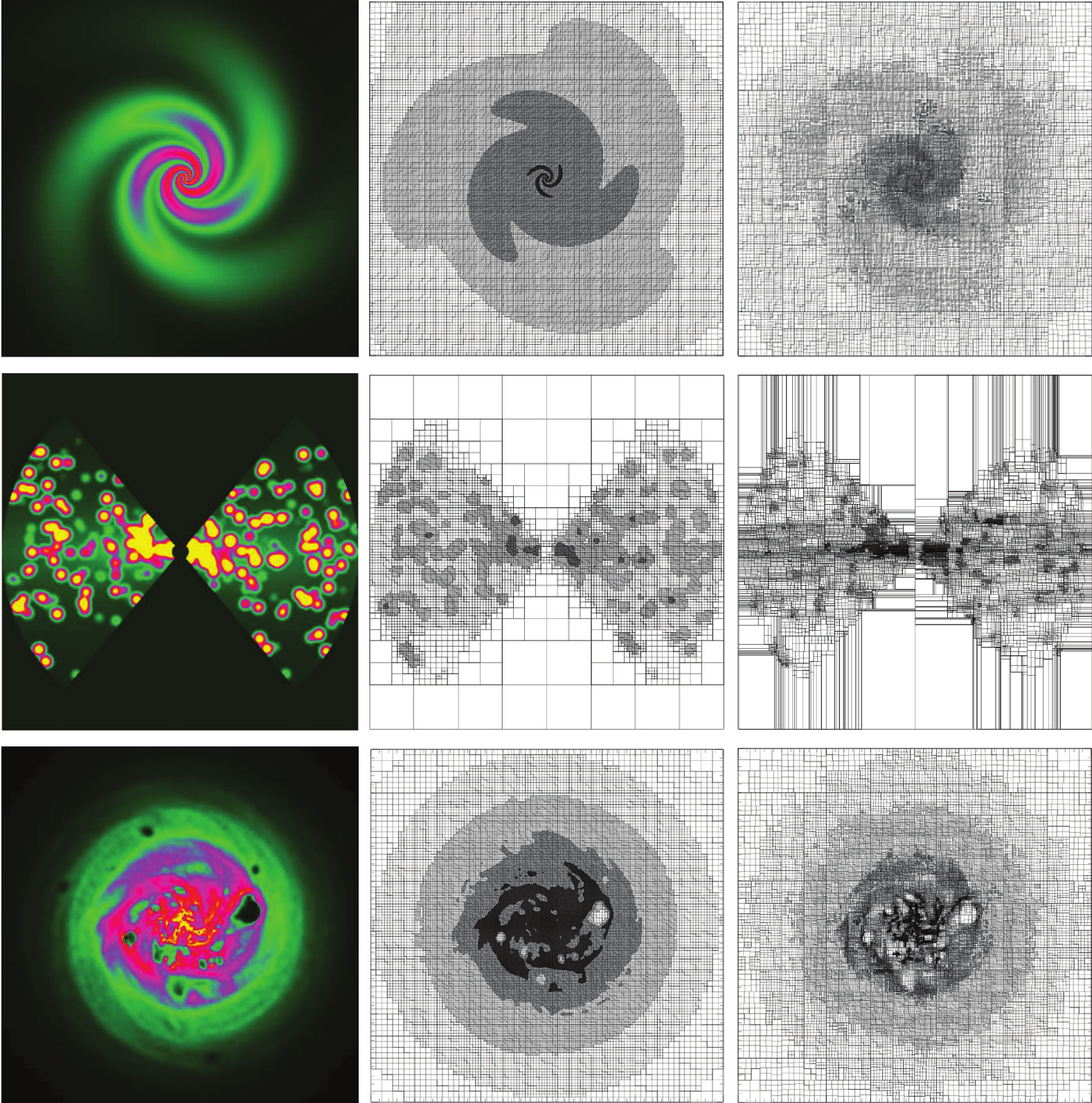}
\caption{Examples of advanced spatial grids used currently for three-dimensional (3D) dust radiative transfer calculations \citep{2013A&A...554A..10S}. The panels are 2D plane cuts through octree-based grids used in Monte Carlo simulations; similar grids are applied in ray-tracing techniques. The left-most panels of each row show the dust density, the central panels the grid distribution in a regular octree structure, and the right-most panels the grid distribution in a barycentric octree structure. The top row represents a regular, analytical model of a double-exponential disc with a three-armed logarithmic spiral density perturbation. The middle row corresponds to a clumpy active galactic nuclei model consisting of several optically thick clumps in a smooth density distribution \citep{2012MNRAS.420.2756S}. The bottom row corresponds to a late-type disc galaxy model resulting from N-body/smoothed-particle hydrodynamics simulations with strong supernova feedback \citep{2012MNRAS.422.2609R}. In all cases, the grids contain between 3 and 4 million cells.}
\label{OcttreeGrids.pdf}
\end{figure}

For many astrophysical applications, the density values cover orders of magnitudes and are highly structured (e.g., a turbulent gas cloud with filaments and shocks). The same is true for sources such as dust grain emission, the distribution of stars, or a layer within a photon dissociation region (PDR) emitting in the MIR. Consequently, the complexity of spatial grids in 3D continuum RT ranges from simple linear Cartesian grids \citep{1991JQSRT..45...47S} to adaptively refined Cartesian grids \citep{2001A&A...379..336K, 2006AA...456....1N, 2012A&A...544A..52L} to multiwavelength adaptive mesh refinement (AMR) grids \citep{2002JQSRT..75..765S}. {\bf{Figure~\ref{OcttreeGrids.pdf}}} illustrates complex dust distributions for three different cases using refined Cartesian grids. In principle, the RTE could be solved on unstructured, dynamic grids such as those used in line RT \citep{2011MNRAS.412..935P, 2010A&A...515A..79P}. Finally, an analytical formula could provide the density or source distribution.

The optimal grid would be based on changes in radiation field intensity. As we do not know the radiation field a priori (this is the goal of the RT calculations), it is extremely challenging to generate such an optimal grid. First, the spatial grid is only 3D, whereas the intensity is defined in 6D (spatial, direction, and wavelength). For example, the intensity in a certain direction can remain constant between two positions, while the intensity in another direction can vary drastically between the same positions. The radiation field also depends on the wavelength, so the optimal spatial grid is different for each wavelength. The dust mass density, or the optical depth, could serve as the starting point on which the grid could be based (see, e.g., {\bf{Figure~\ref{OcttreeGrids.pdf}}}, which shows 3D RT octree grids with a refinement criterion based on the total dust mass in a cell). However, the cellÕs optical depth alone is not suitable to define a grid refinement criterion, as the strength of the radiation field can show strong gradients even in regions where the optical depth is small. In summary, the optimal grid should combine the distribution details of both the dust and source terms such that it captures the variation of the radiation field intensity.

Various approximations are used, given the difficulty in creating an optimal spatial grid. To review the variety of grids and their purposes, it is constructive to distinguish three spatial grid classes. They are used in various combinations by different solvers and in the different astrophysical communities.

\subsubsection{Local mean intensity storage grids}

This most common class of grids stores the mean intensity with the goal of resolving its variation. In the end, the gridÕs resolution determines the spatial resolution of the obtained solution.

To achieve a reasonable sampling of the gradients in the mean intensity, it is possible to calculate a series of spatial grids that are refined using the local optical depth averaged over all directions for each wavelength. The computational effort to calculate such grids is negligible compared to the effort of solving the RTE \citep{2002JQSRT..75..765S}. The drawback of multiwavelength grids is that many interpolations have to be performed to assemble the wavelength-dependent radiation field. Such interpolations are time consuming and introduce interpolation errors. As a compromise, grids optimized for tracking the variation in the local optical depth for the wavelengths that dominate the radiative energy locations can resolve the important regions at the expense of excess grid points. \citet{2006AA...456....1N} proposed an alternative way of building the grid in which an initial calculation defines the explicit locations of several photons, which are then used to refine the grid to have a constant absorption rate in each grid cell.

\subsubsection{Density and source grids}

The grids storing density and source information can originate from a discretization of simple physical models to deep structure grids designed to resolve the kinematic processes. Common examples of the latter are the AMR grid and smoothed-particle (SP) distributions used in hydrodynamical simulations \citep[see, e.g.,][]{2004ApJ...615L.157S}, which typically incorporate several tens of millions of cells and many levels of refinement or smoothed-particle hydrodynamics (SPH) particles. Although RT calculations can be performed directly using such density grids \citep{2002MNRAS.330L..53A}, in most cases the density information is stored on coarser grids to meet the storage and speed limitations of the 6D RT.

\subsubsection{Solution grids}

The third class of grids is designed to calculate the solution directly at the grid points by advancing from one cell to the next in one step. The refinement criterion is defined to minimize solution errors of a certain order \citep{2002JQSRT..73..557S} or to provide a flexible grid that minimizes solution errors due to the coarse spatial coverage of the physical quantities \citep{2010A&A...515A..79P}. These grids are well suited for MHD codes. Solving the RTE directly on a grid accumulates discretization errors causing, in the case of finite-differencing solvers, for example, a smearing of beams. Additional numerical methods or higher-order corrections compensate for these numerical diffusion errors. In some applications, the source function may also contain a smaller number of discrete sources such as stars. These can be considered outside the main spatial grid or by using smaller subgrids \citep{2005AA...439..153S}.

\subsection{Direction Grid}

There are two major challenges in defining a fixed direction grid for 3D RT. First, the radiation field can be strongly peaked due to nearby sources, and a smearing of this beam due to insufficient direction resolution will result in remote regions not being illuminated accurately. Because the optimal intensity discretization can vary greatly across the domain, a globally optimal refinement of direction space is not usually possible. Second, even direction grids optimized to be equally spaced on the unit sphere \citep{1996JQSRT..56...97S} require many points to provide good angular resolution. The HEALPix method, which subdivides the unit sphere in pixels of equal size, offers another possible regular direction grid \citep{2005ApJ...622..759G}. For example, a grid with 600 direction points equally spaced on the unit sphere provides a mean resolution of $\sim$10$^\circ$ only, and it takes 10,000 grid points to achieve a mean resolution of $\sim$2.5$^\circ$. For a protoplanetary disk, this resolution corresponds to a 4-AU hot dust region placed at a distance 100 AU from the central star.

The two main RT solution techniques treat the discretization of directions quite differently. In MC, the direction of the photons is not discretized but sampled from a probability function. For example, the calculation of the scattered intensity in MC is based directly on the scattering phase function, allowing an arbitrarily precise solution. In RayT, a discrete direction grid is used for all calculations. The scattered intensity calculation is done on the direction grid, potentially undersampling the scattering process in the direction space of the solution.

Another issue in direction space is the divergence of rays or photon tracks. The chance to miss physically important parts of the domain increases with the distance from the current point or cell. When tracing the radiation from a single source, this can lead to large errors in the computed radiation field for distant cells, or long runtimes to increase resolution using more photons or rays. Solutions to this issue exist and are discussed in the Monte Carlo Solution Method and Ray-Tracing Solution Method sections, below.

\subsection{Wavelength and Dust Grain Grids}

One must consider variations in source spectra and dust opacity when choosing a wavelength grid. The spectrum of the radiation sources should be covered well enough to resolve the overall shape and any important spectral features (e.g., emission lines). The wavelength grid should resolve variations in the grain properties (e.g., opacities and scattering properties). Where there are features in the dust properties (e.g., 2175\AA\ bump, MIR aromatic/PAH features), the wavelength grid should resolve the feature, ideally including a point at the maximum of each feature as well as enough points to Nyquist sample the feature.

Beside the discretization of the variables entering the intensity directly, the size distribution of the grains needs to be discretized if not given analytically. The grain size discretization can have a strong impact on the radiation field. The extinction of the radiation is the sum of the extinction of the different species, but the emissivity of individual grains is coupled to the intensity of the incoming radiation field (see Section~\ref{sec_RT_dust_emission}).

\section{THE RAY-TRACING SOLUTION METHOD}
\label{raytr}

RayT is a method widely applied in physics and computer graphics to describe the propagation of electromagnetic waves or particles through a medium with varying properties. Important applications outside astrophysics cover the propagation of radio signals in the ionosphere, the investigation of heating by plasma waves, sound waves in the ocean, optical design of lenses, tomographic reconstruction of the EarthÕs interior, and image generation in computer graphics. In an RT context, RayT follows the change of intensity in a particular direction, which is the basic transport problem described by Equation~\ref{RTE-General}. The MC solution technique described in the next section is a sophisticated variant in which a probabilistic approach is taken to choose the direction of the photon package representing the ray. Several 2D continuum RT applications use RayT solvers (see, e.g., the benchmark comparison in \citealt{2009AA...498..967P}), and 3D RayT is based on many techniques first developed in 2D.

In the remainder of Section \ref{raytr}, we describe the basic ingredients and challenges for using RayT as a solver of the general 3D continuum RTE. The solution on a single ray under various numerical and physical conditions is given in Section~{\ref{raytr1}}. The global solution and the treatment of source terms coupling directions are discussed in Section~{\ref{raytr2}}.

\subsection{Ray-Tracing Solution for a Single Ray}
\label{raytr1}  

The elementary RayT operation is to solve the first-order differential (Equation~\ref{RTE-GeneralPath}) within a spatial density grid cell along a given direction. The mass density $\rho_0$, the mass extinction coefficient $\kappa_0$, and the source function $j_0$ are assumed to be constant within the cell for a given wavelength, so that we can determine the intensity $I(s + \Delta s, \lambda)$ from $I(s,\lambda)$ using Equation~{\ref{RTE-FormalSolution}}:
\begin{equation} 
\label{onecell}
I(s+\Delta s, \lambda) 
=
I(s, \lambda)\, \txe^{-\tau_0(\lambda)}
+
\frac{j_0(\lambda)\Delta s}{\tau_0(\lambda)}
\left( 1-\txe^{-\tau_0(\lambda)} \right)
\end{equation}
where $\tau_0(\lambda) = \kappa_0(\lambda)\,\rho_0\,\Delta s$. For a ray crossing several cells, the numerical task is to determine the cells that are hit by the ray, calculate the intersection points with the cell borders, and use Equation~{\ref{onecell}} to calculate the change in intensity along the ray in each cell. The first step can be time-consuming, given that an adaptively refined grid is often stored as a tree, and neighbor-search calculations are required. The error in the intensity is defined solely by the finite resolution of the underlying spatial grid.

For clarity in the notations used, we note that the solution of the RTE along rays with constant direction is also called the method of short or long characteristics. Long characteristics refers to calculating the radiation field along a ray through the entire computational domain. Because several rays can cross a certain cell causing redundant calculations, precalculated local column density or optical depth values can be used (short characteristics) to interpolate the contributions along a ray. Usually, a sweep of the ordering of the grid is required to ensure that, for a given direction, all information about positions before the currently considered point are given before performing the step described by Equation~{\ref{onecell}}. This method is less accurate than long characteristics because of the accumulation of interpolation errors. For combined applications of RT and MHD, hybrid methods have been proposed combining the advantages of short and long characteristics \citep{2006A&A...452..907R}.

\subsubsection{Beyond the spatial grid resolution}

There are cases where the best RayT solution is not performed at the density grid resolution. First, the spatial resolution of the density from an AMR MHD calculation with many refinement levels can be too fine for an RT calculation to be done in a reasonable time. One way to deal with this is to interpolate the density to a coarser grid, but some of the information on the physical structure obtained in the AMR run will be lost. Second, the density grid shows strong gradients. One way to soften the gradients is to interpolate the density to a finer grid, but this can still leave the density changes abrupt. Instead, an interpolation scheme can be used to find the density in between grid cells. Finally, the density is given analytically, and no step-size information is available. In practice, analytical density structures are not automatically simpler than AMR density grids. The density used by \citet{2010A&A...511A...9S}, for example, for the RT modeling of Spitzer images of the molecular cloud L183 involved 100 3D clumps with Gaussian profiles and 700 free parameters.

The simplest approach to solve the RTE in these cases is to apply an upwinding first-order finite-difference scheme. The relation between two intensities located at $s$ and $s + \Delta s$ along the ray then is
\begin{equation} 
\label{FD}
I(s+\Delta s,\lambda)
=
I(s,\lambda)
\left[1-\tau(s,\Delta s,\lambda) 
\right]
+ 
j(s,\lambda)\,\Delta s\,,
\end{equation}
using the local optical depth $\tau(s,\Delta s,\lambda) = \kappa(s,\lambda)\,\rho(s)\,\Delta s$. The step size $\Delta s$ is chosen to be small enough to stay in the optically thin limit, allowing the Taylor expansion of the exponential to be truncated after the first term ($\exp[-\tau] \approx 1-\tau$). The advantage of this scheme is that it is fast; the disadvantage is that first-order errors can accumulate along the ray. Moreover, the steps become very small in regions of high optical depth, although the radiation field takes a simple form in this limit.

Trial steps can improve the accuracy. Using a Runge-Kutta scheme, a step with the size $\Delta s/2$ can be made to calculate a second estimate of the derivative. The scheme then advances to the next point by using a linear combination of both derivatives. The linear factors are chosen after comparing them with the Taylor series of $I(s)$, then letting the first-order term vanish.

This improvement can be repeated to achieve a better solution, at the cost of repeated calculations or lookup of the density and source terms. \citet{2002nrc..book.....P} propose an advanced ray tracer based on fifth-order Runge-Kutta accuracy. It is coupled with an adaptive step-size control using an embedded form of the fourth-order Runge-Kutta formula. Values determined by \citet{1990ACMTMS...16..201C} are used as parameters for the error truncation. The tracer steps are chosen adaptively and with full error control. Such a tracer is therefore the first choice for astrophysical RT problems with moderate optical depth variations that are not solved at the resolution of the density grid. This explicit error control is an important characteristic of the RayT solution technique.

\subsubsection{High optical depths} 

There are 3D RT applications with strong gradients in the source function or in the local optical depth. In particular, high optical depth $\tau\gg1$ occurs in all star formation regions as well as in AGN tori. All algorithms used to solve the RT problem encounter problems in correctly describing the intensity changes in the optically thick case. Adaptive grid RT solution techniques refine the regions into too many cells. Only moment methods that are applied in conjunction with MHD solvers are posed well to treat the optically thick regime, but they in turn fail to describe a strongly peaked radiation field arising in low optical depth regions.

To illustrate the difficulties of high optical depths for the RayT method, we assume a simple modified blackbody thermal source term for a single dust species $j(s,\lambda) = \kappa(s,\lambda)\,\rho(s)\,B(s,\lambda)$ in Equation~{\ref{FD}}, which then becomes
\begin{equation} 
\label{hightau}
I(s+\Delta s,\lambda)
-
I(s,\lambda)
=
\tau(s,\lambda) 
\left[ B(s,\lambda)-I(s,\lambda)
\right].
\end{equation}
When the ray moves from one point to the next, most of the radiation at $s$ is damped, as is the source contribution along the path, so that the local radiation at $s + \Delta s$ is dominated by the source contribution. Hence $B-I$ is small, suggesting that the solver can perform large steps. But $B-I$ is multiplied by the large optical depth $\tau$ , amplifying any change in the source term that arises from the spatial variation in the temperature. Thus, to meet accuracy requirements, the tracer must perform small steps.

The second reason to modify the ray tracer is the two exponential functions entering the radiation equation. First, the exponential containing the optical depth has to be calculated precisely along the ray. Given a limited computational range of the computer, an optical depth of 1,000 usually exceeds these limits and makes it necessary to renormalize the intensity. Second, the Planck function rises sharply with $\lambda$ for temperatures in the Wien part $T < hc /\lambda k$.

To solve the two-exponential problem, \citet{2006ApJ...645..920S} proposed using the transformation to the relative intensity $D = I/(I + B)$. For a vanishing source function, it approaches unity; for the optically thick part, where $I(s,\lambda) \approx B(s,\lambda)$, $D(s,\lambda)$ has a value close to $1/2$. The authors showed that the transformation avoids numerical problems caused by the exponentials and that the intensity error amplification by the transformation is less than a factor of 2. As criterion for the use of an approximate solver in the optically thick region $D(s,\lambda) \approx 1/2$, they derived
\begin{equation}
 \frac{|B(s_2,\lambda)-B(s_1,\lambda)|}{B(s_1,\lambda)}
 \frac{1}{\tau(s_1,\lambda)}
 <
\varepsilon
\end{equation}
where $\varepsilon$ is a positive limit for the relative difference $|B-I|/B$. The condition is fulfilled by either small relative changes in the source term or a large local optical depth. In a precalculation along the ray, the regions in which the condition is fulfilled can be identified. Then, a fast solution is obtained for these regions by performing large steps and assuming $D(\lambda,x) = 1/2$. Applying the scheme to a massive molecular cloud core illuminated by a nearby star, \citet{2006ApJ...645..920S} verified speed-up factors of several hundreds compared to those of fourth-order Runge-Kutta solvers.

\subsection{Ray location and global solution of the RTE}
\label{raytr2}  

\begin{figure}
\centering
\includegraphics[width=\textwidth]{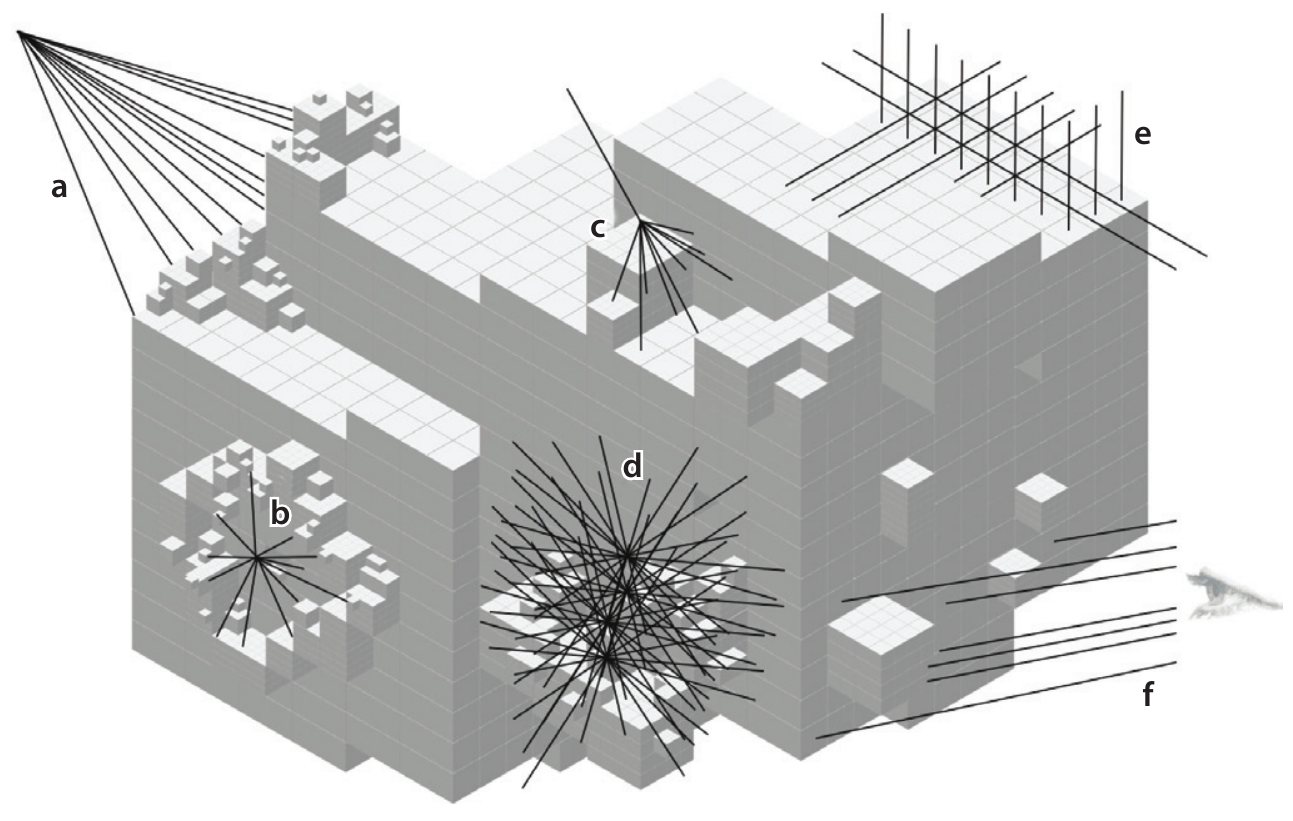}
\caption{Illustration of the various types of rays used in three-dimensional dust radiative transfer calculations performed within an adaptively refined density grid. Only a small fraction of the typical number of rays are shown for ({\em{a}}) an outer radiation source, ({\em{b}}) an inner radiation source, ({\em{c}}) a scattering event, ({\em{d}}) a region with an optical depth near 1, ({\em{e}}) a coarse regular outer grid, and ({\em{f}}) rays to the observer.}
\label{RayT}
\end{figure}

After discussing the solution methods for a single ray through the computational domain, the next step is to determine how to place the rays to ensure that the radiation field is correctly calculated. This is a critical part of the solution process.main.

\subsubsection{Thermal emission}

We describe how to place the rays to calculate the intensity of the radiation field when the source term is dominated by thermal emission from dust grains. {\bf{Figure~{\ref{RayT}}}} illustrates the various ray patterns used in RayT. In what follows, Òplacing a rayÓ describes the basic RayT step that includes defining the ray by a point in space and a direction, solving the intensity along the ray as previously described, and storing the absorbed energy in the cells that are crossed. According to Equation~\ref{onecell}, the intensity of the absorbed radiation per cell is
\begin{equation} 
\label{onecellloss}
I(s,\Delta s,\lambda) 
 =
I(s, \lambda) \left( 1-\txe^{-\tau_0(\lambda)} \right)
+
j_0(\lambda)\,\Delta s
\left[
1 - \frac{1}{\tau_0(\lambda)}
\left( 1-\txe^{-\tau_0(\lambda)} \right)
\right].
\end{equation}
Its contribution to the local radiation field then is calculated from the formula for the mean intensity (Equation~\ref{meanintensity}) and the balance equation (Equation~\ref{balance}). In each RayT step, the local source term (Equation~{\ref{dustemissivity}}) contains the thermal energy of the currently crossed cell; this is updated with each ray crossing.

In RayT, precalculation steps are done to analyze the specific RT problem and accelerate the computations. The optical depth affects the transport within the domain, the thermalization of the radiation, and the appearance of the object on the $\tau(\lambda) \approx 1$ surface for the observer. Therefore a penetration depth analysis is performed at all wavelengths, which determines the optical depth distribution with respect to all discrete sources and the observer \citep[see, e.g.,][]{2002JQSRT..75..765S}. For this calculation, rays are placed from the sources to all grid cells ({\bf{Figure~{\ref{RayT}}a,b}}). In addition, we calculate the optical depth from the source to the cell using
\begin{equation}
\label{tausou}
\tau_{\text{sou}}(s_1,s_2,\lambda)
=
\int_{s_1}^{s_2}
\kappa(s,\lambda)\,\rho(s)\,\txd s
=
\sum_{i=1}^{N_c} \kappa_i(\lambda)\,\rho_i\,\Delta s_i,
\end{equation}
with the ray crossing $N_c$ cells with their individual $\kappa_i$, $\rho_i$, and crossing length $\Delta s_i$. $\tau_{\text{sou}} \approx 1$ marks the region where most of the source radiation is reprocessed.

The solver also calculates the optical depth from each cell to the observer $\tau_{\text{obs}}$ as described above ({\bf{Figure~{\ref{RayT}}f}}). This information helps us understand which regions are shielded at which wavelengths from the observer, and to calculate the final images once the main RayT has been performed.

The second precalculation is to propagate the initially deposited discrete source energy and the radiation field at the domain boundary through the domain. The calculation is done on a regular grid of rays such as that shown in {\bf{Figure~{\ref{RayT}}e}}. This calculation also determines the optical depth between cells, $\tau_{\text{cc}}$, on a coarse spatial grid at all wavelengths, therefore providing information on which regions in the domain exchange significant amounts of radiation. In star formation applications, for example, regions often do not connect significantly at UV wavelengths.

If the source of radiation is the interstellar radiation field, the precalculation can be very time-consuming, as it comprises rays from all directions to all cells at all wavelengths for which the optical depth can reach 1. However, there are many cases where this precalculation is less time- consuming. For example, when the dust properties are constant in the domain, the optical depth is $\tau_{\text{sou}} = \kappa\,N_{\text{col}}$, and this is an integral over the density only.

The main RayT calculations are performed separately for each wavelength. Rays are placed through the grid points in the $\tau_{\text{sou}} \approx 1$ layers covering all directions. In this way, additional resolution is provided where the largest changes in the radiation field are expected. Moreover, additional rays from discrete sources are placed to help resolve the illumination of the $\tau_{\text{sou}} \approx 1$ layers. This can be important, for example, in the case of an accretion disk atmosphere that is illuminated by a protostar through a narrow solid angle. The order in which the computation is done for each wavelength influences the convergence. In RayT, information has already been transported from the sources to the cells during the precalculation step. Therefore, starting with wavelengths covering the peak of the re-emitted photon energy (e.g., in the MIR) can speed up the information transport in thermally dominated problems. For problems dominated by low optical depths in the UV/optical, the order is less important as the dust self-heating is small compared to the dust heating from UV/optical photons.

In practice, the placing of rays is controlled by the maximal number of rays per cell $N_l$ and the width of the $\tau_{\text{sou}} \approx 1$ layers: $\Delta \tau_l$, where $|\tau_{\text{sou}} - 1| < \Delta\tau_l$.The placing and iteration over wavelength is stopped when the change in the energy deposited in all cells drops below a chosen change limit and the mean field distribution has been determined. If the energy does not converge after placing $N_l$ rays in all cells within the $\tau_{\text{sou}} \approx 1$ layers, a second run with all rays is performed, using the updated energy information from the first run. If still no convergence is obtained in this run, $N_l$ is increased to improve the resolution. The number of rays for typical 3D dust RT applications can easily exceed $10^6$.

The final step is to calculate the radiation that is received by the observer from each cell. This calculation makes use of the precalculated $\tau_{\text{obs}}$. Moreover, radiation received directly from the inner and outer discrete sources as well as from the background radiation that is inside the field of view is calculated by placing rays from the sources and the background to the observer.

To illustrate how the rays are placed in various situations, we present a few simple examples. The first is a molecular cloud core with a gas mass of 1 $M_\odot$ that is illuminated by a strong MIR radiation field that dominates its thermal budget. The core has no internal source of radiation except the thermally emitting dust; furthermore, the self-heating of the dust by neighboring dust grains within the core can be neglected. The precalculations will not yield any $\tau$ surfaces, given that the optical depth is too low in the MIR radiation field. The main calculation will therefore be to propagate the external anisotropic radiation field through the core with little need for refinement, by placing additional rays. Correspondingly, the rays can be placed dense enough in direction space to resolve all features of the extended illumination source \citep[e.g., a nearby PDR;
see][]{2005A&A...434..167S}.

The second example is a binary star surrounded by a circumstellar disk. The precalculation will identify the inner disk and the disk surface as the $\tau\approx1$ zone for the NIR wavelengths. The ray pattern will therefore be dense at the inner rim, and in the zone above and below the disk. At MIR or longer wavelengths, the ray pattern will be less dense in the atmosphere, given that the $\tau\approx1$ range will move deeper into the disk, as the thermally emitting inner dust is important. This application has an additional complication in that it requires calculating the scattered light (see below).

A central source illuminating more complex structures is a problem for all RT solvers: A proper resolution requires many spatial cells and therefore many rays from the star to the cell and from the cell into the surrounding regions. For RayT, if a fixed direction grid is used, more and more cells are overlooked when the diverging rays reach the outer regions. For such applications, beam splitting can be used to split a ray in order to sample several neighboring cells \citep{2002MNRAS.330L..53A}.

\subsubsection{Including scattered radiation}

Whereas the thermal source contribution can be calculated using the mean intensity $J(\bfx,\lambda)$ per cell, the scattered light intensity depends on the direction of the incoming radiation, the optical depth for scattering, and the phase function of the grains. The RayT scheme that includes thermal reradiation and scattering must therefore store J and the intensity and direction of the incoming radiation for each cell and wavelength.

Every RayT step is expanded to include directly scattered light by calculating the amount of radiation that is scattered toward the observer using the precalculated $\tau_{\text{obs}}$. In many applications, this singly scattered radiation is a good fraction of the total scattered light in the computational domain. Including multiple scattering in RayT methods significantly increases the computational requirements. Applications using 3D dust RT based on RayT with scattering sources have been presented for massive disk candidates \cite{2006A&A...456.1013S} and grain growth in molecular cloud cores \cite{2010A&A...511A...9S}. Additional codes using RayT solvers and alternative techniques to deal with high optical depth are described in \citet{2009AA...498..967P} and the references within. Fortunately, there are specific astrophysical regimes in which multiple scattering is not crucial or reasonable approximations can be made to limit the computational requirements.

The standard ISM dust grain distribution can be approximated as a mean grain with a 0.1-$\mu$m grain size. Such grains have very low scattering efficiency in the MIR and, thus, scattering is often neglected if calculations are done at MIR or longer wavelengths. However, caution should be exercised, as coagulated grains in dense molecular clouds scattering can be be important even at MIR wavelengths \citep{2010A&A...511A...9S}. In addition, dust grains have a strongly forward-peaked scattering phase function, and a portion of the scattered light can be transported along the already calculated rays \citep[see, e.g.,][]{2003AA...401..405S}. For the wavelengths in which dust shows a more isotropic scattering phase function (e.g., MIR), additional rays can be added to carry the scattered radiation. Adding scattering is fairly straightforward in, for example, modeling star formation regions. Models of these regions will have a high number of rays in cells that are identified using the $\tau\approx1$ search algorithm, allowing scattering to be calculated along existing rays. Finally, the high albedo of dust grains and the distribution of the scattered intensity over the unit sphere naturally reduce the intensity of the scattered radiation with subsequent scatterings allowing an intensity threshold to be used to limit the number of scatterings needing to be calculated.

In astrophysical objects where the optical depth is very small, or very large, or where the $\tau\approx1$ layer is not resolved, calculating multiple scatterings can be ignored. If $\tau\ll1$ for an object (e.g., the diffuse ISM or the zodiacal light), then single scattering completely dominates the scattered intensity, and multiple scattering can be ignored. For $\tau\gg1$, the main modeling goal is usually to calculate the thermal dust emission. Because scattering is generally important at optical and shorter wavelengths, most of the energy at these wavelengths is absorbed in the $\tau\approx1$ regions of the object. Thus, not calculating the scattered light properly will not have a large impact on the dust emission results. For very embedded sources such as massive young stellar objects, the stellar energy is completely converted to radiation at MIR or longer wavelengths before it leaves the core, and therefore scattering does not influence the outer radiation field. Finally, if a model does not resolve optically thick regions (e.g., in some extragalactic applications) and thus does not have $\tau\approx1$ regions, then all the rays can be used to determine the global field and to transport the singly scattered radiation.

\subsection{Ray-Tracing Error Analysis}

Machine precision limits the precision of a solution for the RTE along a ray, through a spatial grid on which the opacity and source terms are described analytically. This can be done only with machine precision errors. For example, the previously mentioned Runge-Kutta solver with adaptive step-size control provides a good compromise between accuracy and computational cost, while providing explicit error control.

The main source of error for solutions on a single ray is the interpolation error of the density and source function from the underlying grid. It can accumulate if the density grid shows strong (and partially unresolved) gradients. The precalculation of $\tau\approx1$ regions and shielded areas helps to characterize how well the chosen grid describes the underlying physical problem; refining the grid with this information can reduce the interpolation errors in the intensity distribution. Unlike solvers that calculate the intensity on a grid (e.g., finite-difference solvers), RayT solvers create no diffusive errors (i.e., beam smearing).

There is no general formula for the global error in the achieved intensity distribution, because deviations caused by coarse resolution due to the placement of rays are hard to quantify. A good test for the global accuracy of the thermal conversion of radiation is to calculate the energy output of the source term integrated over the domain, and to compare it to the energy in the radiation leaving the domain. Another test for the overall resolution of the important regions in the domain is based on the number of crossing rays per cell. All the cells that contribute significantly to the radiation field overall should be crossed by many rays. Having important cells with minimal rays is an indication that the grid was underresolved.

A practical test to understand and measure the global error is to increase the spatial or wavelength resolution by inserting more rays or adding wavelength grid points, and testing the stability of the solution.

\section{THE MONTE CARLO SOLUTION METHOD}
\label{monte}

The MC method is a general computational technique that is used widely in many different fields, including numerical mathematics, physical sciences, finance, and medicine. Its name reflects a variety of stochastic or probabilistic techniques, which all have in common that they solve equations by sampling random numbers. The MC methods are particularly interesting when applied to transport systems, which was the motivation for their first application in the 1940s. For a general overview of MC methods as a tool for transport problems, see, e.g., \citet{Dupree2002}, \citet{Kalos2009} or\citet{2011BASI...39..101W}. Its application to dust RT problems in an astrophysical context has a history of more than 40 years \citep[see, e.g.,][]{1970A&A.....9...53M, 1974ApJ...190...67R, 1974AJ.....79..948W, 1977ApJS...35....1W}. In the past four decades, it has become a mainstream method for 3D dust RT calculations.

The basis of MC RT is to treat the radiation field as the flow of a large but finite number of photon packages (often called photons). Each individual photon is followed along its journey through the dusty medium. At every stage in its journey, the characteristics that determine the path of each photon (such as its birth location, initial propagation direction, or the distance along the path until the next interaction with a dust grain) are determined in a probabilistic way by generating random numbers from an appropriate probability density functions (PDF). At the end of the simulation, the radiation field is recovered from a statistical analysis of the photon paths. Hence, the MC technique simulates the RT instead of explicitly solving the RTE.

Central to all MC techniques is the process of generating random numbers from a given PDF $p(x)\,{\text{d}}x$. Thus an algorithm is needed that returns a set of numbers $X$ such that the probability that $X$ lies in the infinitesimal interval between $x$ and $x + {\text{d}}x$ is equal to $p(x)\,{\text{d}}x$. The starting point for such algorithms, which are key to the MC process, is a (pseudo)random number generator. This is a code that generates uniform deviates (random numbers with an equal probability to be chosen in the unit interval between 0 and 1). To generate random numbers from another, arbitrary PDF, one almost always applies appropriate operations on one or more uniform deviates. The most popular methods are the so-called transformation and rejection methods, details of which can be found in \citet{Devroye1986} or \citet[][Ch.~7]{2002nrc..book.....P}.

In the remainder of Section~\ref{monte}, we gradually introduce the ingredients and techniques that are combined to develop a 3D MC dust RT code. We start in Section~\ref{SimpleMCRT.sec} by describing the simplest MC RT technique, as it is the basis for all modern MC RT techniques. These simple techniques are sufficient for MC calculations in geometries with large degrees of symmetry, such as 1D spherical or slab geometries. For 3D applications, however, they would result in very inefficient codes. Fortunately, there are various weighting schemes that improve the performance of this technique, making modern MC RT quite efficient. Some of these weighting schemes have a strong heritage in RayT methods, making most modern MCRT codes hybrids between the classical MC and RayT techniques. Several of these weighting schemes were developed for 2D geometries, especially cylindrical geometries. The use of weighting schemes is critical for 3D MCRT; in Section~{\ref{WeightedMCRT.sec}}, we discuss several of these techniques.

\subsection{Simple MC RT}
\label{SimpleMCRT.sec}

The simplest MC calculation consists of considering the RT problem at a single wavelength $\lambda$. We consider a source of photons, characterized by the source term $j_*(\bfx,\bfn,\lambda)$, and a distribution of dust, characterized by the dust density $\rho(\bfx)$. Throughout the calculation, the state of each photon is tabulated by its energy, position, direction of travel, and polarization state. For 3D RT, the Cartesian coordinate system is usually used, resulting in the photonÕs position being stored as $\bfx=(x,y,z)$ and the direction using direction cosines as $\bfn=(n_x,n_y,n_z)$. The polarization state is stored using the Stokes vector $\bfS = (I,Q,U,V)$.

 \subsubsection{Step 1: birth.} 
 
The first step in a photonÕs life cycle is its birth, i.e., its injection into the computational domain. If $N$ is the number of photons in the model run and $L_{\text{tot}}(\lambda)$ is the total luminosity of the source, the luminosity carried by each photon is $L = L_{\text{tot}}(\lambda)/N$ . The initial position $\bfx$ and propagation direction $\bfn$ are to be chosen randomly according to the source term
$j_*(\bfx,\bfn,\lambda)$, which means that they need to be sampled from the PDF
\begin{equation}
  p(\bfx,\bfn)\,\txd\bfx\,\txd\bfn
  =
  \frac{j_*(\bfx,\bfn,\lambda)\,\txd\bfx\,\txd\bfn}
  {\iint j_*(\bfx,\bfn,\lambda)\,\txd\bfx\,\txd\bfn}
  =
  \frac{j_*(\bfx,\bfn,\lambda)\,\txd\bfx\,\txd\bfn}
  {L_{\text{tot}}(\lambda)}
\end{equation}
In many cases, for example, for emission by stars or thermal emission by dust grains, the emission is isotropic, which implies that the initial propagation direction can be chosen randomly from the unit sphere:
\begin{equation}
  p(\bfn)\,\txd\bfn 
  = 
  \frac{\txd\bfn}{4\pi}
  =
  \frac{\sin\theta\,\txd\theta\,\txd\phi}{4\pi}
\end{equation}
Generating random values for $\theta$ and $\phi$ from this distribution can be done easily using the transformation method. In other cases, for example, when we deal with external emission illuminating an interstellar cloud or anisotropic emission from an AGN accretion disc, different PDFs for the initial propagation direction need to be considered \citep[see, e.g.,][]{2003AA...399..703N, 2012MNRAS.420.2756S}. For most cases, the photon is assumed to be emitted unpolarized; i.e., $\bfS = (1,0,0,0)$.

\subsubsection{Step 2: determination of the interaction point.} 

Once the photon is launched into the dusty medium, the next step consists of randomly determining whether it will interact with a dust grain, and if so, where this interaction will take place. The PDF that describes the free path length before an interaction is most conveniently described in optical depth space, where it has an exponential distribution, $p(\tau)\,\txd\tau = \txe^{-\tau}\,\txd\tau$. The optical depth $\tau$ to which a particular photon travels along its path before it interacts with a dust grain, is drawn from this exponential distribution. This is done easily using the transformation method: We simply pick a uniform deviate $\xi$ and solve the equation
\begin{equation}
  \xi = \int_0^\tau \txe^{-\tau'}\,\txd\tau'
\end{equation}
for $\tau$. Integrating and solving gives $\tau = -\ln\xi$ (recall that the distributions of $\xi$ and $1-\xi$ are equivalent). If $\tau$ is greater than the optical depth $\tau_{\text{path}}$ to the surface of the system in the direction the photon is traveling, the photon escapes the system, and the life cycle of this particular photon is over. Otherwise, the photon will interact with the dust medium at the location along the path corresponding to the traveled optical depth $\tau$.

The next step is converting the traveled optical depth $\tau$ to a physical path length $s$, such that we can move the photon to the interaction site. This means that we have to integrate along the path and solve the integral equation
\begin{equation}
  \int_0^s \kappa_{\text{ext}}(s',\lambda)\,\rho(s')\,\txd s' = \tau
\label{solvefors}
\end{equation}
for the path length s. Comparing this equation with Equation~{\ref{tausou}} highlights the intimate link between the MC solution technique and the RayT technique: A large fraction of the calculations in MC simulations are pure RayT operations.

In practice, MC codes virtually always use a grid structure of dust cells on which the dust density and the optical properties are discretized. The integral in Equation~{\ref{solvefors}} then reduces to a sum over the consecutive grid cells along the path, and the integral equation comes down to summing the optical depth along the path until $\tau$ is reached. This calculation is often one of the more computationally intensive portions of MC RT and is a strong driver for choosing a grid optimized for the particular astrophysical object of interest. For a Cartesian model grid, the distance traveled in each cell is easy to calculate, as is the next grid cell along the path. For hierarchical grids, more complex neighbor-search algorithms may be required.

\subsubsection{Step 3: absorption and scattering.} 

Once the path length $s$ has been calculated, the photon moves from its old location $\bfx$ to its updated location, i.e., the interaction site $\bfx + s\,\bfn$. At this location, the photon can either be absorbed or scattered; the appropriate PDF is hence not a continuous but a discrete function with only two possible values. The probability that the interaction is a scattering event is equal to the dust albedo $a = \kappa_{\text{sca}}/\kappa_{\text{ext}}$. Using a uniform deviate $\xi$, the nature of the interaction is easily determined: If $\xi\leq a$, we have a scattering event and if $\xi>a$ an absorption event.

In the case of an absorption event, this is the end of the photonÕs life cycle. If dust emission is to be calculated in the simulation, the absorbed photon luminosity is stored in the interaction cell. This absorbed luminosity will be used at a later stage to calculate the dust emission spectrum, which can then be used as the source function for another MC cycle.

If the interaction is a scattering event, the next step consists of determining the new propagation direction. In the case of isotropic scattering, this just comes down to generating a new random point from the unit sphere. In the case of anisotropic scattering, the new propagation direction $\bfn$ is chosen according to the PDF
\begin{equation}
  p(\bfn)\,\txd\bfn 
  =
  \frac{\Phi(\bfn,\bfn',\bfx,\lambda)\,\txd\bfn}{4\pi}\,,
\end{equation}
where $\Phi(\bfn,\bfn',\bfx,\lambda)$ is the scattering phase function, and $\bfn'$ is the original propagation direction before the scattering event. For spherical dust grains, the scattering phase function is dependent only on the scattering angle between $\bfn$ and $\bfn'$. For the HG phase function, the most popular approximation to the real phase function, the generation of a random scattering angle and hence the calculation of the new propagation direction can be done analytically \citep{1977ApJS...35....1W}. For polarized RT, the scattering process is more complex, as the scattering phase function is dependent on the polarization of the photon and, in addition, scattering changes the polarization state of the photon. A detailed description of scattering events in the case of polarized radiation can be found in, for example, \citet{1994AA...284..187F}; \citet{1995ApJ...441..400C}; or \citet{2007AA...465..129G}.

With its new propagation direction determined, the photon can continue its journey through the dusty medium. This means that the second and third steps are repeated until the photon either escapes from the system or is absorbed by a dust grain. (Cases in which the optical depth is very large and the number of scatterings is correspondingly large are possible. As a result, it is common to impose a limit on the number of scatterings to calculate. Consequently, the high optical depths are not sampled well, and there are potential systematic errors in the calculation. For most cases, the systematic errors from imposing a maximum number of scatterings are negligible.) This entire process is repeated for all the photons until the last photon has left the dusty medium. The RT at different wavelengths can be independently calculated given there is no explicit wavelength coupling in the RT. There is an implicit coupling when dust emission is included in the modeling. In this case, the absorbed luminosity at every wavelength is stored in every dust cell on the spatial grid. After finishing the simulation at all the wavelengths, the absorbed luminosity is used to compute the mean intensity $J(\bfx,\lambda)$ of the radiation field and subsequently the dust source term $j_\txd(\bfx,\lambda)$. It is common to include the dust emission as a second source of photons by first computing the RT from the primary sources, computing the RT for the dust emission, recomputing the dust emission with the new radiation field, and iterating until a set convergence level is reached.

In simple MC RT, the output desired from the calculation is usually the view of the system from a particular observer location. Images of the system are constructed by projecting the position of each photon that escapes in the direction of the observer within some angular tolerance onto the plane of the sky. This plane is divided into pixels, and the images are built from those photons. Many of the photons will escape the system in directions other than the observer and are not counted, unless symmetries (e.g., circular or cylindrical) can be exploited. Special care must be applied when constructing the images that give the polarization state of the scattered flux, specifically the Q and U images. As part of the construction of these images, the polarization vector of the photon must be rotated so that it is referenced correctly in the image plane.

\subsection{Weighted MC RT}
\label{WeightedMCRT.sec}

\begin{figure}
\centering
\includegraphics[width=0.85\textwidth]{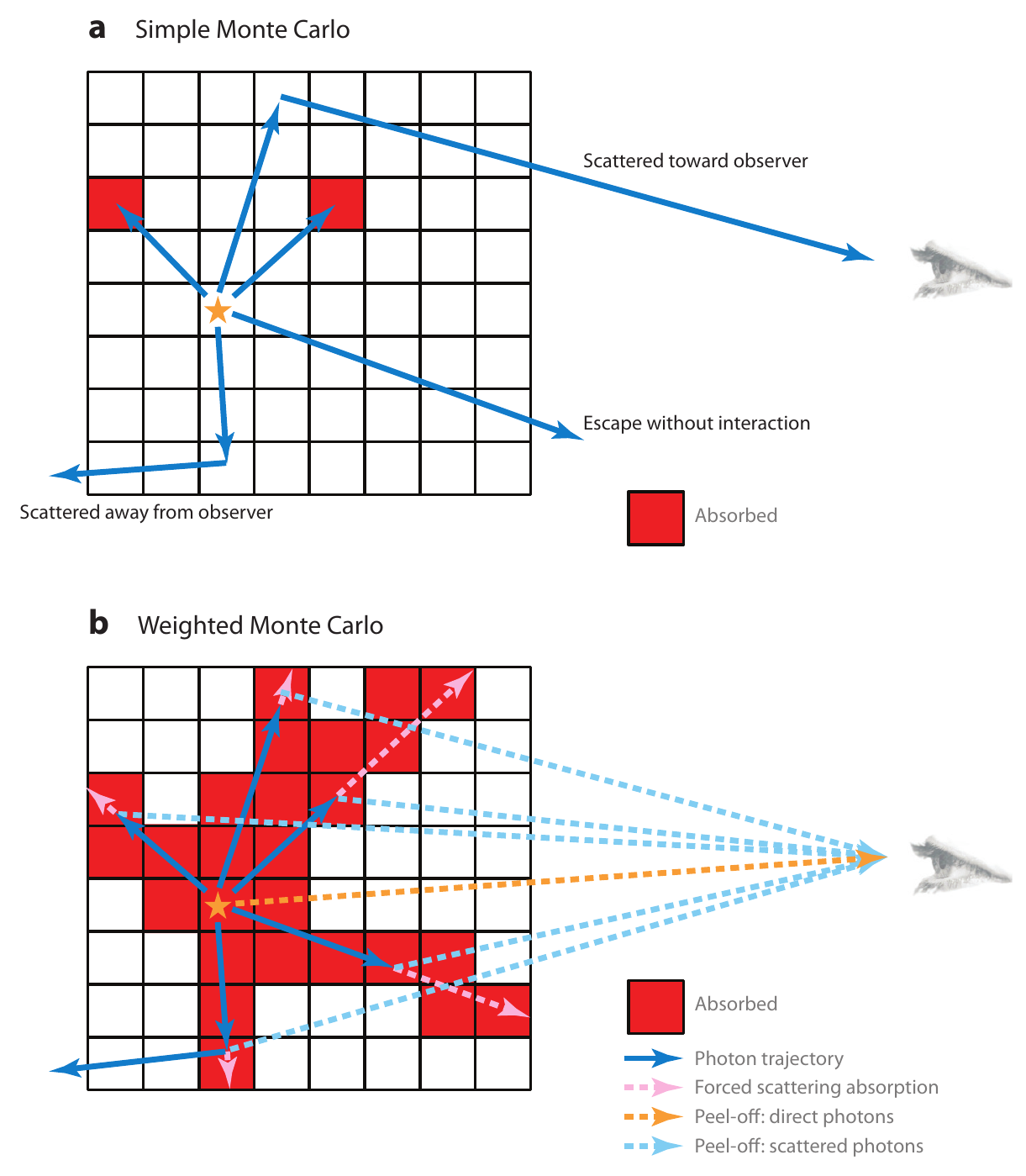}
\caption{({\em{a}}) Simple and ({\em{b}}) weighted Monte Carlo (MC) simulations are illustrated graphically. This example includes five photons resulting in one scattering photon reaching the observer and two absorption events for simple Monte Carlo radiative transfer (MC RT). This same set of five photons produces five scattered photons toward the observer and 28 absorption events for weighted MC. The improvement in computational efficiency of weighted MC is seen clearly.}
\label{MonteCarlo.pdf}
\end{figure}

Simple MC RT is the easiest to understand, but it would be extremely inefficient for full 3D calculations. For 1D and 2D MC RT calculations, there are symmetries, such as spherical and cylindrical, that can be exploited, and thus the number of photons needed to achieve accurate results is relatively small. In the case of 3D dust RT, there are no symmetries by definition; this is one of the motivations for several acceleration methods. Most of these acceleration methods are well established and validated, whereas others still have a more experimental character. Almost all of these methods were first developed for 2D RT calculations. {\bf{Figure~\ref{MonteCarlo.pdf}}} illustrates the benefits of weighted MC techniques as compared to simple MC RT methods.

The basis of all acceleration methods is to assign a weight $W$ to each photon and modify this weight. The weight of each photon is equivalent to the fraction of the luminosity of the emission source carried by that photon, i.e., the number of photons in each photon package. Several acceleration techniques use the idea of biasing, i.e., generating random numbers from a PDF $q(x)\,\txd x$ rather than from the appropriate PDF $p(x)\,\txd x$. This biased behavior is corrected for by assigning the weight $W = p(x)/q(x)$ to the photon. The biasing technique is used in many MC applications and can be a very effective way of reducing the variance \citep{Dupree2002}. It is, common practice, for example, in MC numerical integration, where it is known as importance sampling.

\subsubsection{Biased emission.} 

A direct application of the biasing technique is the so-called biased emission. The initial propagation direction of the photons launched into the dust medium is usually determined from the angular part of the source term $j(\bfx,\bfn,\lambda)$. There are cases, however, where an increased level of emission in particular directions is desired, for example, to increase the signal-to-noise ratio in particularly interesting directions, such as the polar regions of a star with an accretion disk \citep{1984ApJ...278..186Y}. In this case, the emission of photons is biased toward the directions of interest, and the initial weight of the photon is determined using the standard biasing weight formula. The same technique can also be applied to the spatial part of the source term to increase the number of photons emitted from regions with a low emission rate \citep{2005AA...440..531J}.

This technique of biased emission has the potential to strongly increase the efficiency of an MC simulation. However, it has the drawback that it is very model-dependent, and therefore requires significant manual interaction. It is, in a sense, comparable to the placement of the rays in the RayT method.

\subsubsection{Absorption-scattering split.} 

This acceleration method allows for a photon to contribute to both absorption and scattering at each interaction site. Instead of randomly choosing the nature of the interaction, the photon is split in two parts: one that is absorbed and one that scatters. The fraction absorbed is equal to $(1-a)$ times the current weight of the photon. For the scattered part that continues its life cycle through the dust, the weight is multiplied by a factor $a$.

\subsubsection{Forced scattering} 

Instead of having each photon either scatter or escape the system, the photons can be forced to scatter each time \citep{Cashwell1959}. In the simple MC routine, the randomly generated optical depth is compared to the total optical $\tau_{\text{path}}$ along the photonÕs path. This approach is problematic for regions with low optical depth: In those regions, many photons just leave the system without interacting with the dust, resulting in low efficiency of dust scattering and heating. A way to avoid this low efficiency is the technique of forced scattering, which limits the values of the randomly chosen optical depths to the range between 0 and $\tau_{\text{path}}$. This can be achieved by biasing the PDF from which the optical depth is generated. Instead of sampling from the actual exponential PDF, we consider an exponential distribution cut off at $\tau = \tau_{\text{path}}$. Properly normalized, this PDF reads
\begin{equation}
  q(\tau)\,\txd\tau
  =
  \begin{cases}
    \;\dfrac{\txe^{-\tau}\,\txd\tau}{1 - \txe^{-\tau_{\text{path}}}} 
    &\qquad \tau<\tau_{\text{path}}, \\
    \;0
    &\qquad \tau \geq \tau_{\text{path}}.
  \end{cases}
\end{equation}
Generating a random $\tau$ from this distribution is straightforward and guaranteed to produce an interaction before the photon has left the system. The compensation to be applied to the weight of the photon is a factor
\begin{equation}
  W_{\text{fs}}
  =
  \frac{q(\tau)}{p(\tau)}
  =
  1 - \txe^{-\tau_{\text{path}}} \, . 
\end{equation}
One issue with forcing every scattering, when combined with the absorption-scattering split, is that there is no longer a natural stopping criterion for the scattering calculation. In the original MC cycle, photons end their journey when they are either absorbed by the dust or leave the dusty system. The common solution is to set a minimum weight for a photon, below which the photonÕs life cycle is terminated. The value of this termination weight is usually set to be very low, after the equivalent of many scatterings. An alternative solution is to force only the first scattering and revert to the standard scattering calculation for subsequent scatterings.

\subsubsection{Peel-off technique.} 

For 3D cases, it is usually desired to calculate the appearance of a system for an observer at a particular orientation. Simple MC RT is particularly inefficient in building up such an image as only the photons that are emitted from the system in the direction of the observer contribute to the output appearance. In addition, some blurring of the image is inherent as photons that are within some tolerance of the desired direction are used. This inefficiency can be eliminated by requiring that all photons directly contribute to the output images by calculating the portion of the photon that is emitted from sources and scattered at every interaction point in the observerÕs direction \citep{1984ApJ...278..186Y}. The weight factor of a photon in the direction of the observer is
\begin{equation}
  W_{\text{po}} 
  = 
  p(\bfn_{\text{obs}})\,\txe^{-\tau_{\text{obs}}}
\end{equation}
where $\tau_{\text{obs}}$ is the optical depth from the position of the emission or scattering event, and $p(\bfn_{\text{obs}})$ is the probability that the photon will be directed toward the observer. For example, for isotropic emission, $p(\bfn_{\text{obs}}) = 1$, and after a scattering event we have $p(\bfn_{\text{obs}}) = \Phi(\bfn,\bfn_{\text{obs}},\bfx,\lambda)$.

The peel-off technique is a pure application of RayT, again highlighting the connection between both approaches. Peel off is probably the most important acceleration technique for 3D MC RT simulation. It has a significant computational cost, however, as it requires a calculation of the optical depth from the emission or scattering location toward the observer after every emission or scattering event. One way to alleviate this computational burden is by precalculating the optical depth toward the observer for each cell. Another possibility is to store the information for each cell and create the images at the end of the simulation \citep[see, e.g.,][]{2000AA...360.1187D, 2006AA...459..797P, 2009AA...497..155M}. Using a precomputed $\tau_{\text{obs}}$ or stored information to compute the image does come with a price: loss of subgrid resolution in the model images. The benefit of this approximation in decreased computation time has to be weighted against the loss of subgrid resolution for the particular modeled astrophysical object.

\subsubsection{Continuous absorption.} 

The accuracy of the re-emission of energy absorbed by the dust can be enhanced by absorbing not just at the interaction site, but also along the path the photon travels. Depending on the implementation, this can be done up to the location of the scattering event \citep{1999AA...344..282L}, or along the entire path (\citealt{2003AA...399..703N}, \citealt{2011ApJS..196...22B}). In the latter scenario, the photon is effectively split in $N + 2$ different parts: one part $W_{\text{esc}}$ that leaves the system (and is hence not accounted for anymore), one part $W_{\text{sca}}$ that is scattered at the interaction location (determined stochastically), and $N$ parts $W_{\text{abs},j}$ that represent the fraction absorbed in each of the $N$ cells along the photonÕs path. These different fractions are, respectively,
\begin{equation}
  W_{\text{esc}} = \txe^{-\tau_{\text{path}}}\, ,
\end{equation}
\begin{equation}
  W_{\text{sca}} = a \left(1-\txe^{-\tau_{\text{path}}}\right) ,
\end{equation}
and
\begin{equation}
  W_{\text{abs},j} = (1-a) \left(\txe^{-\tau_{j-1}}-e^{-\tau_j}\right) ,
\end{equation}
where $\tau_j$ is the optical depth measured from the photonÕs location to the surface of the $j$th cell along the path. An alternative interpretation of this continuous absorption approach is that it estimates the mean intensity of the radiation field, not by means of the absorbed luminosity, but by means of counting the luminosity that passes through each cell. The strength of the technique lies in the fact that all photons contribute to the calculation of the absorption rate of each cell they pass through, and not only of those cells with which they interact. This is particularly useful for the optically thin regime, which has very few absorptions in the simple MC approach.

\subsubsection{Instantaneous dust emission.} 

The traditional method to computing the self-consistent dust emission in an MC RT simulation consists of running an independent MC simulation at every individual wavelength and storing the absorbed luminosity over the computational grid. In a second stage, the dust emission spectrum is calculated in each dust cell and used as a secondary source term. This approach inevitably leads to iteration, as the dust emission itself affects the radiation field. It is possible to compute the output dust emission spectrum from a multiwavelength model without iterating. The instantaneous dust emission technique or frequency distribution adjustment technique emits dust thermal photons immediately after each absorption event in a specific grid cell, with the wavelength of the emitted thermal dust photon carefully chosen to retain thermal equilibrium (\citealt{2001ApJ...554..615B};
\citealt{2005NewA...10..523B}).

Besides eliminating the need for iteration in the computation of the dust emission spectrum, the instantaneous dust emission technique also has the advantage that photon packages are emitted at the exact position where the absorption event took place, so subgrid resolution is achieved. One disadvantage is that the absorption/re-emission event takes place in a single cell, which results in a poor convergence rate of the radiation field in cells with a low dust absorption rate. This makes the classical iterative technique with continuous absorption more efficient for 3D simulations than the instantaneous dust emission technique, at least when applied in its original form \citep[e.g.,][]{2009ApJ...690.1432C}. This problem can be alleviated by applying a combination of the instantaneous dust emission and continuous absorption techniques. In this hybrid method, the photon packages are followed through the domain using the instantaneous dust emission technique, but the final dust emission spectrum of the cells, used to create images and SEDs, is calculated based on the continuous absorption approach \citep{2006AA...459..797P}. A second problem for the instantaneous dust emission technique is that it was originally designed to work with equilibrium dust emission. Several ways have been explored to adapt the instantaneous dust emission technique for transiently heated grains (\citealt{Krugel2007}, \citealt{2008ApJ...688.1118W}, \citealt{2012ApJ...751...27H}).

\subsubsection{High optical depths.} 

In regions of high optical depth, the simple MC routine becomes very inefficient, as photons can be trapped in a virtually never-ending loop of scattering events. This problem is largely solved when the absorption-scattering split is applied, as the weight of the photon then decreases with every scattering event terminating when a very low weight is reached. However, in regions with extreme optical depths (such as in the midplane of circumstellar discs around protostars) or at wavelengths where scattering largely dominates absorption (such as the far-UV), this can still imply a significant computational burden. A solution to this problem is to mirror or reflect photons from high optical depth regions and use the diffusion approximation to find the RT solution in these regions. The diffusion approximation allows for multiple interaction steps to be calculated in a single computation. An elegant solution that is well adapted to the MC method is to solve for the RT in optically thick cells using a modified random walk technique that also uses the diffusion approximation \citep{2009AA...497..155M, 2010AA...520A..70R}.

\subsubsection{Polychromatism.} 

For multiwavelength RT, it is possible to significantly speed up the calculation by considering photon packages that consist of photons of all wavelengths. The advantage of this technique is that an MC run is simultaneously solved at all wavelengths, instead of a run for each wavelength. The difficulty in this approach is that many of the PDFs that describe the life cycle of a photon are wavelength dependent, such as the path length distribution or the scattering phase function. One solution is to consider partly polychromatic photon packages, which shift to monochromaticism as soon as wavelength-dependent PDFs are involved \citep{2005AIPC..761...27B}. A more advanced option is to perform the calculations at one reference wavelength $\lambda_{\text{ref}}$ and use the biasing technique to adjust for the wavelength-dependent PDFs (\citealt{2005AA...440..531J}; \citealt{2006MNRAS.372....2J}). The efficiency gains of full polychromatic RT are large given that every photon calculated can contribute to the output images and radiation field density at all wavelengths instead of a single wavelength. But there is a known significant complication -- the biasing factors can be very large and, as a result, can dominate the results at a particular wavelength. This has a systematic effect on the results that is hard to control and therefore, this method should be considered experimental.

\subsection{Uncertainties for Monte Carlo}
\label{sec_unc}

In the simple (unweighted) MC RT, the noise in the output quantities (i.e., scattered intensity, polarization, etc.) scales as $N^{-1/2}$, where $N$ is the number of photons. In the case of weighted MC, the uncertainties in the output quantities do not scale directly with $N^{-1/2}$.

The uncertainties can be calculated by using the dispersion in the average properties of the photons used to determine an integrated quantity \citep{2001ApJ...551..269G}. If the integrated quantity is $X$, then
\begin{equation}
  X = \sum_{j=1}^N x_j = N\,\overline{x}\, ,
\end{equation}
where $x_j$ is the contribution of the $j$th photon to $X$, $N$ is the total number of photons in the model run, and $\bar{x}$ is the average contribution each photon makes to $X$. The uncertainty in $X$ is then $\sigma_X = X (\sigma_x/\bar{x})$, where $\sigma_x$ is the standard deviation of $\bar{x}$, calculated using
\begin{equation}
  \sigma_x^2 
  = 
  \frac{1}{N(N-1)}\sum_{j=1}^N (x_j - \overline{x})^2 
  = 
  \frac{1}{N - 1}
  \left( \overline{x^2} - \overline{x}^2 \right).
\label{eq_unc}
\end{equation}
The equations for the uncertainties in output quantities given above can be used to explicitly enforce a particular uncertainty level in the final results of a model run. The uncertainties in the output quantities of interest can be checked during the model run and the number of photons adjusted dynamically to achieve the desired accuracy. For example, when calculating a multiwavelength model, more photons can be added at the wavelengths with higher optical depths compared to wavelengths with lower optical depths. As the number of photons is determined from the statistics of the particular run itself, it automatically takes into account the full RT solution, including the locations of the photon sources, dust, etc. The output quantities to be used for uncertainty control can range from the global flux, to pixels in resolved images, to the radiation field density in each cell. Additionally, it is possible to improve the convergence to the needed accuracy by identifying areas of the model with high noise (e.g., particular cells with few photon interactions) and dynamically adding photons to those areas (e.g., through biased emission). Dynamically determined uncertainties to improve the convergence of a model have been used in limited cases; this is clearly an area for future improvement.

Uncertainties associated with the model setup itself are more difficult to quantify. Evaluating the systematic uncertainties due to specific model choices is usually done by trying other parametrizations or increasing the grid resolution and evaluating how the output quantities change. Example model setup choices that are prone to systematic uncertainties are the dust density grid, specific parametrizations of the photon sources, and the wavelength grid. These kinds of uncertainties are clearly the most difficult to diagnose and generally rely on the expertise of the coder and user.

\section{CHALLENGES IN MODELING OBSERVATIONS}
\label{advan}

The uses of 3D dust RT codes are many. They range from helping us understand the impact of locally clumpy dust on dust RT (\citealt{1996ApJ...463..681W, 2000ApJ...528..799W}, \citealt{2008AA...490..461B}), to modeling observed images of objects to derive the source and dust distributions \citep{2012MNRAS.419..895D, 1994ApJ...432..641G, 2005A&A...434..167S}, to predicting the appearance of an object that has been modeled with an HD code \citep{2004ApJ...615L.157S, 2004ApJ...610..801B,
  2010MNRAS.403...17J, 2011AA...536A..79R}, to investigating the ensemble behavior of objects to derive average source and dust properties \citep{2011ApJ...738..124L}, and to testing observability of objects for instrument and observation planning \citep{2008Ap&SS.313..109W, 2012A&A...547A..58G}. To illustrate the challenges in modeling observations, we focus on the modeling of objects to derive their physical properties from observational data. The common approach is to perform some steps of the following scheme:
\begin{enumerate}
\item Choose a parameterized model for the dust density, radiation sources, and dust optical properties.
\item Discretize the problem with the choice of grid and use the RT code to derive an SED and/or images for this object model.
\item Compare simulated and observed SEDs or images.
\item Evaluate the differences in the SEDs or images and change the model parameters and/or the model assumptions to minimize these differences. Then, repeat step 2.
\item Find the parameter sets that come closest to describing the observed data.
\item Evaluate the observational and theoretical errors that are important for the comparison.
\end{enumerate}
Although the scheme seems straightforward, the full set of steps has rarely been carried out for existing, published 3D dust RT models, as each of these steps provides challenges for 3D dust RT.

\subsection{Model Choice} 

\begin{figure}
\centering
\includegraphics[width=\textwidth]{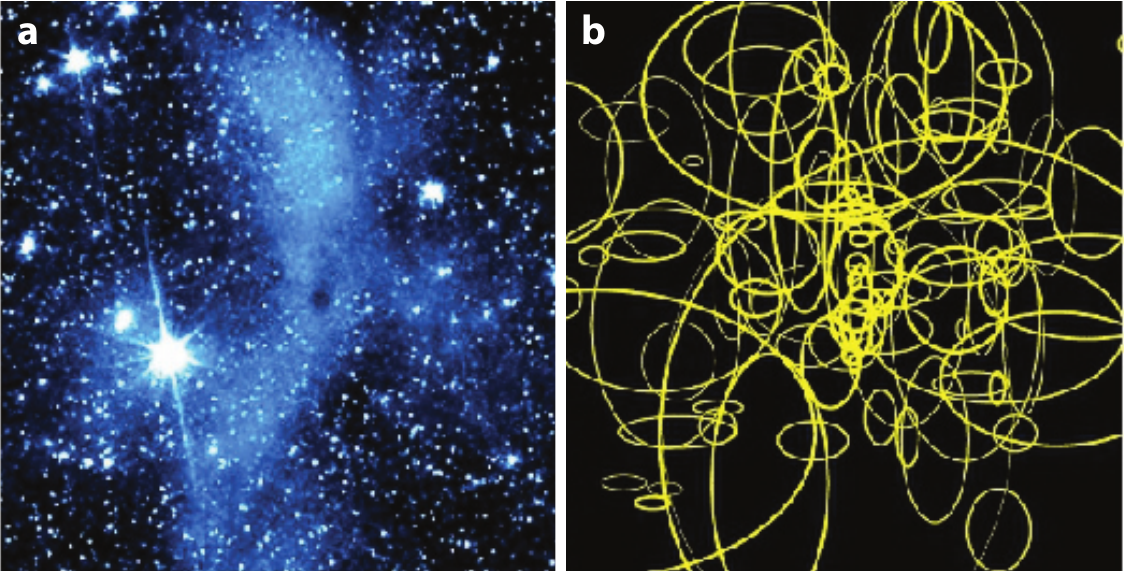}
\caption{({\em{a}}) {\em{Spitzer Space Telescope}} image of the molecular cloud L183 at 3.6 $\mu$m revealing "coreshine", which is scattered light from the densest part of the cloud. ({\em{b}}) Spatial modeling based on basis functions that have Gaussian density structures in all three coordinates \citep{2010A&A...511A...9S}. The ellipses give the full width at half maximum of the various Gaussians in the plane of sky.}
\label{Coreshine.pdf}
\end{figure}

\begin{figure}
\centering
\includegraphics[width=\textwidth]{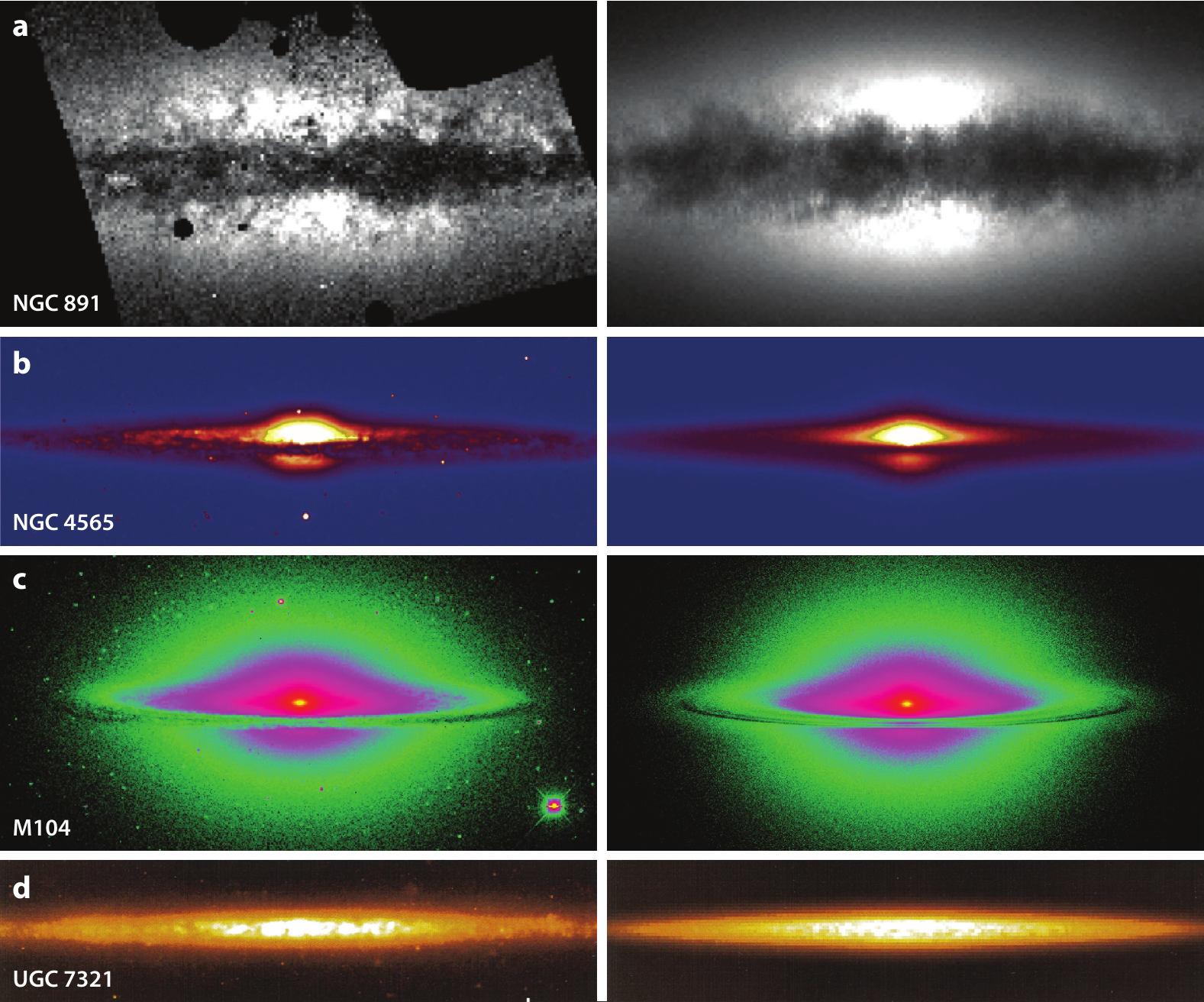}
\caption{Examples illustrating the current state of the art in fitting dust radiative transfer models to images of galaxies. The left panels show the observed images; the corresponding panels on the right are the fits to these images. ({\em{a}}) A clumpy three-dimensional (3D) spiral galaxy model fit to a {\em{Hubble Space Telescope}} B-band image of the prototypical edge-on spiral galaxy NGC\,891 \citep{2012ApJ...746...70S}. ({\em{b}}) A 2D disc galaxy model fit to a Sloan Digital Sky Survey $g$-band image of NGC\,4565 \citep{2012MNRAS.427.2797D} using a fully automatic fitting based on genetic algorithms \citep{2013A&A...550A..74D}. ({\em{c}}) A detailed 2D model for the Sombrero galaxy, fit to a V-band image \citep{2012MNRAS.419..895D}. ({\em{d}}) A 3D clumpy disc model fit to an R-band image of the edge-on low surface brightness galaxy UGC 7321 \citep{2001ApJ...548..150M}.}
\label{GalaxiesFit.pdf}
\end{figure}

The aim is to choose an appropriate model that allows meaningful physical information to be derived from the given observational data. This is a general topic, but 3D RT modeling has special features that are important to consider. The data are usually SEDs and/or images, and we can investigate the number of independent information bits a priori, for example, by counting the number of wavelengths for which the flux density values have been measured. This is then compared to the number of model parameters.

Most 3D models include 10 to several hundreds of free parameters, depending on the complexity of the spatial model and the assumed dust properties. Even for attempting to reproduce only the global SED of an object, it is difficult to stay below 10 parameters, given that modeling a dusty 3D structure requires a few length scales and/or power laws to describe the density, dust properties, and viewing angles. {\bf{Figure~{\ref{Coreshine.pdf}}}} illustrates the large number of free parameters needed to reproduce the 3D dust distribution of a molecular cloud. {\bf{Figure~{\ref{GalaxiesFit.pdf}}}} shows four examples of state-of-the-art RT fits to dusty galaxy images. Such RT model fits are designed to determine the parameters that describe the intrinsic 3D distribution of stars and dust in both large-scale geometric parameters (e.g., stellar and dust scale lengths and heights) and measures of the small-scale inhomogenity or clumpiness of the dust.

Starting with more free model parameters than data points is often considered to produce meaningless parameter values. However, a careful exploration of the parameter space can detect ambiguities and reveal redundant parameters. As long as the parameter space exploration can be afforded computationally, starting with a detailed model and analyzing the parameter ambiguities is usually the most accurate way to model an object.

\subsection{Gridding}

As discussed in Section~\ref{setup}, there are a variety of options for discretizing space, direction, wavelength, and the dust model. If the grid fails to resolve structures that are important at a particular wavelength, the corresponding model image will be inaccurate. A good example is the surface layer of an accretion disk illuminated by a central star. If the layer is not resolved well, the description of the radiation field that is scattered in the layer will be poor and the scattered light images inaccurate. Therefore, the discretization step often includes running the RT code before the start of the modeling to verify that expected features are present in the image or SED and that changes/refinement do not influence the overall appearance of the object.

\subsection{Comparison of Models and Data}

Once the model images or SEDs have been calculated, the results should be convolved with the beam of the observed instruments/telescopes, and the sampling should be made equal (pixel size for images). It is important to apply detection limits, especially when studying faint structures. For interferometric observations, the incomplete coverage of all spatial scales implies that the comparison should be performed in the (u,v) plane rather than the spatial plane to achieve the highest fidelity. Calibration uncertainties and correlated noise properties in the observations should be understood and included in the model, as opposed to observations comparison alone. In addition, foreground emission (e.g., caused by the zodiacal light) should be carefully removed or modeled, especially for on-off chopping observing modes that remove this emission as part of the data reduction process. Generally, modelers can benefit from good communication with experienced observers. Another potential source of error in comparing model results and data can be a displacement of structures due to observational uncertainties. Additional translation parameters can correct this and provide physical insight by deriving improved positioning.

\subsection{Exploration of the Parameter Space}

The parameter space of 3D dust distributions is large: A uniform coverage of a 10D parameter space with five grid points in each parameter, would imply $\sim$10 million individual RT calculations. Therefore, almost all 3D RT modeling of data has been done "by hand", that is, starting from a point in the parameter space and then varying just one parameter to explore the variations in the results. This drastically reduces the coverage in parameter space at the expense of exploring correlations between parameters. In structures with strong gradients, inhomogeneously distributed radiation sources or varying optical depth, the resulting radiation field may react strongly to changes in the model parameter. In this case, low-coverage Òby handÓ explorations are likely quite unreliable.

To evaluate the model and the variation in the model parameters, the difference between the model data and the observed data must be defined (often using a $\chi^2$ metric), and the fitting procedure must then be to minimize this difference by varying the parameters. Some standard optimization methods applied in astrophysics are gradient descent, NewtonÕs method, Metropolis optimization, and genetic algorithms. The latter two are able to leave local minima with the goal of ending in the deepest minimum providing the best-fitting model parameters.

There are advancements in the application of automated fitting techniques that are already in use for 2D RT calculations and are starting to be used for 3D RT calculations as well. One such technique is the Metropolis algorithm that has been applied to 3D dust RT in the form of simulated annealing \citep{2005A&A...434..167S}. Precomputing a large grid of models is a promising technique. \citet{2007ApJS..169..328R} used a parametrized circumstellar disk model to create a large grid of precalculated SEDs for a 2D configuration and then used the grid to fit the disk parameters and to characterize uncertainties in fit parameters. Other promising techniques are 2D RT fitting techniques based on the Levenberg-Marquardt algorithm \citep[see, e.g.,][]{1999A&A...344..868X}, the downhill-simplex method \citep{2007A&A...471..765B}, and genetic algorithms \citep{2013A&A...550A..74D}. Some of these techniques have already been applied to 3D structures, albeit in limited parameter spaces (see, e.g., \citealt{2000ApJ...528..799W}; \citealt{2012ApJ...746...70S}).

Quantifying the ambiguity of the derived parameters is also important. The 2D circumstellar disk modeling of SEDs in \citet{2007ApJS..169..328R} provides a template for determining the ambiguity when a complex model is applied to only a few SED points. A general strategy to assess ambiguity is to explore the variation of the fitting metric (e.g., $\chi^2$) in the vicinity of the best-fit parameters and characterize the overall degree of variation by random or grid-based parameter space exploration.

\subsection{Error Analysis}

Each of the RT solvers has its own sources of errors due to approximations, the underlying grid, undetected regions of the computational domains, etc. Tracing these errors, measuring them, and estimating their importance is nontrivial. RayT provides error control for the solution along one ray, and MC can provide error control on the number of photons needed; however, whether the rays are sent through the right parts of the domain is not well quantified. For modeling, a comparison of these errors with the solution and observational errors determines the sensitivity of the observed data to the specific science question of interest.

\subsection{Inverse RT}

The true challenge of RT modeling is to invert it, i.e., determine the 3D density structure of the dust and the sources, and the dust properties from a set of images taken at different wavelengths. Direct inversion is computationally impossible in 3D with current or expected computing capabilities. In 1D, an analytic inversion can be performed under special assumptions \citep{2002JQSRT..74..183S}, and numerical 1D inverse RT modeling was used in \citet{2010MNRAS.406.1190D}. A forward RT method using many iterations provides a good solution but is limited to fairly simple RT applications. A prominent difficulty is the loss of information due to the line-of-sight integration inherent in the observations. Multiwavelength observations can help disentangle the line-of-sight integration. For example, points in the center of a molecular core are better shielded and cold compared with points in the outer parts, so millimeter observations can be used to constrain the core center and shorter wavelength observations to constrain the outer parts. Such molecular cores are simple enough to enable inverse RT to be done. For example, \citet{2005A&A...434..167S} used a 3D background radiation field illuminating a 3D model core and fitted the MIR and millimeter images of the core $\rho$ Oph D, deriving the density and temperature structure with assumed dust properties.

\section{CODES AND BENCHMARKS}
\label{codes}

\subsection{Available 3D codes}

\begin{table*}
\centering
\scriptsize
\caption{List of published three-dimensional dust radiative transfer (RT) codes used previously and currently for astronomical applications. The codes are ordered alphabetically. Given that many codes are first used for specific investigations with full code details then published at a later date (references in table), it is difficult to establish true code development dates. Note on Type: Finite-differencing (FD), Monte Carlo (MC), or ray-tracing (RayT).}
\label{listofcodes}
\vspace*{1em}  
\begin{tabular}{|l|l|l|l|}
\hline
Code name & Type & Reference & Main application \\ \hline 
SKIRT & MC & \citet{2003MNRAS.343.1081B, 2011ApJS..196...22B} & Galaxies, AGN \\
(no name) & MC & \citet{2004ApJ...610..801B, 2007ApJ...663.1055B} & Star-forming (SF) clouds \\
TRADING & MC & \citet{1996ApJ...465..127B}, \citet{2008AA...490..461B} & Galaxies\\
RADISHE & MC & \citet{2007ApJ...658..840C, 2009ApJ...690.1432C} & Galaxies \\
(no name) & MC & \citet{2005MNRAS.362..737D} & SF clouds \\
RADMC-3D & MC & Dullemond (Manusc.\ in prep.) & SF disks \\
MOCASSIN & MC & \citet{2005MNRAS.362.1038E} & Photoionized regions \\
(no name) & MC & \citet{1994AA...284..187F} & SF disks \\
(no name) & MC & \citet{2004AA...415..617G} & SF clouds \\ 
STOKES & MC & \citet{2007AA...465..129G} & AGN \\
DIRTY & MC & \citet{2001ApJ...551..269G, 2001ApJ...551..277M} & Galaxies, nebulae\\
TORUS & MC & \citet{2000MNRAS.315..722H, 2004MNRAS.350..565H} & SF disks \\
(no name) & MC & \citet{2012ApJ...751...27H} & SF disks, AGNs \\
SUNRISE & MC & \citet{2006MNRAS.372....2J, 2010MNRAS.403...17J} & Galaxies \\
CRT & MC & \citet{2003AA...397..201J}, \citet{2005AA...440..531J} & SF clouds \\
(no name) & MC & \citet{1999AA...344..282L, 2005AA...429...19L} & supernovae \\ 
MCMax & MC & \citet{2009AA...497..155M, 2011Icar..212..416M} & SF disks \\
STSH & MC & \citet{2008AA...490..673M} & SF disks \\ 
MCTRANSF & MC & \citet{2003AA...399..703N}, \citet{2006AA...456....1N} & SF disks \\
mcsim mpi & MC & \citet{2006AA...445.1015O} & Carbon stars \\
MCFOST & MC & \citet{2006AA...459..797P} & SF disks \\
HYPERION & MC & \citet{2011AA...536A..79R} & SF clouds \\
PHAETHON & MC & \citet{2004AA...420.1009S, 2005AA...439..159S} & SF cores \\  
STEINRAY & FD & \citet{2003AA...401..405S} & SF disks\\
    & RayT & \citet{2006ApJ...645..920S} & SF cores \\
(no name) & FD & \citet{1991JQSRT..45...47S} & SF disks, AGNs \\
HO-CHUNK & MC & \citet{2002ApJ...574..205W} & SF disks\\
MC3D & MC & \citet{2000CoPhC.132..166W}, \citet{2003CoPhC.150...99W} & SF disks, SF cores, AGNs\\
(no name) & MC & \citet{1999ApJ...525..799W}, \citet{2001ApJ...554..615B} & SF disks, galaxies \\
(no name) & RayT & \citet{1997AA...325..135X}, \citet{2000AA...353..117M} & Galaxies \\
    \hline
  \end{tabular}
\end{table*}

Until the mid-1990s, few RT codes could handle the absorption, scattering, and thermal emission by interstellar dust in general 3D geometries. The available codes were either limited to 1D or 2D geometries, or were not able to calculate the full 3D problem (e.g., missing either scattering or thermal emission). The spectacular increase in computational capabilities during the past two decades, as well as the development of more powerful techniques to solve the RT problem, has led to the creation of many 3D RT codes. To our knowledge, there are almost 30 operational codes that can handle the full dust RT problem, i.e., including absorption and multiple anisotropic scattering in a general 3D geometry. {\bf{Table~{\ref{listofcodes}}}} gives a list of published 3D dust RT codes that have been or are being used for astronomical applications. The application fields of the different codes vary widely, ranging from prestellar cores and circumstellar discs to AGN and galaxies.

With many different codes and several techniques available, it might be difficult for a potential user or future developer to decide which one to choose. The choice for a given code or solution technique should be driven primarily by the specific nature of the problem being solved. Although most of the 3D codes in {\bf{Table~{\ref{listofcodes}}}} are applicable for a wide range of RT problems, virtually all of them have been developed with a particular application in mind, and hence have been optimized for that particular goal.

The programming language can also influence the choice; most existing 3D RT codes have been coded in FORTRAN, C, or C++. There are no features in the 3D RT problem itself or the two solution methods presented in this review that make certain languages preferential to others. The main driver is speed: Because the 3D RT problem is computationally very challenging, all codes should be developed in a language suitable for high-performance computing. Moreover, parallelization is becoming increasingly important as a means to increase computational speed and memory availability. Several codes are designed to work on shared-memory or distributed-memory clusters and/or adopt graphical processing units \citep[e.g.,][]{2006MNRAS.372....2J, 2010NewA...15..509J, 2011ApJS..196...22B, 2011AA...536A..79R, 2012ApJ...751...27H}.

The choice of a given code or method primarily depends on the specific needs of the application and/or the personal preferences of the user or developer. Nevertheless, the different approaches have some general strengths and weaknesses. These need to be considered as rough guidelines only, as there are significant differences among codes based on the same approach. For example, simple MC codes based on the naive techniques explained in Section~{\ref{SimpleMCRT.sec}} are many orders of magnitude less efficient than modern MC codes that use weighting schemes.

Generally speaking, setting up a reasonably efficient 3D MC RT code can be done in the time taken to do a PhD. Most MC RT codes are intrinsically 3D codes, and hence the shift from 1D or 2D codes to a full 3D geometry is fairly straightforward. For RayT codes, however, the increase in complexity when moving from 1D and 2D to 3D is much steeper. These differences explain the relative scarcity of general 3D RayT codes compared with the MC codes in {\bf{Table~{\ref{listofcodes}}}}; a similar comparison of 1D or 2D codes would result in a table with a much larger fraction of RayT codes. A big part of this complexity lies in the placement of the rays, which needs manual adjustment in RayT codes but is done automatically in MC codes. The RayT precalculation step needed for the manual placement of rays reveals an advantage for the MC method, as it identifies the critical locations in the model that dominate the appearance of the object. Other advantages of the MC method are that it needs less storage than RayT codes when scattering is included, and that it is widely used with many coders in the community improving its application to astrophysics with new algorithms, whereas RayT methods are mainly improved outside astrophysics. One strength of the RayT codes is the stronger and explicit error control. For example, the precalculation step can identify areas that are potentially under-resolved, allowing for changes to the grid to provide adequate resolution for the full calculation. At this point, MHD codes tend to use RayT rather than MC techniques \citep[e.g.,][]{2006A&A...448..731H, 2012A&A...537A.122K}.

\subsection{Benchmark efforts}

\begin{figure}
\includegraphics[width=\textwidth]{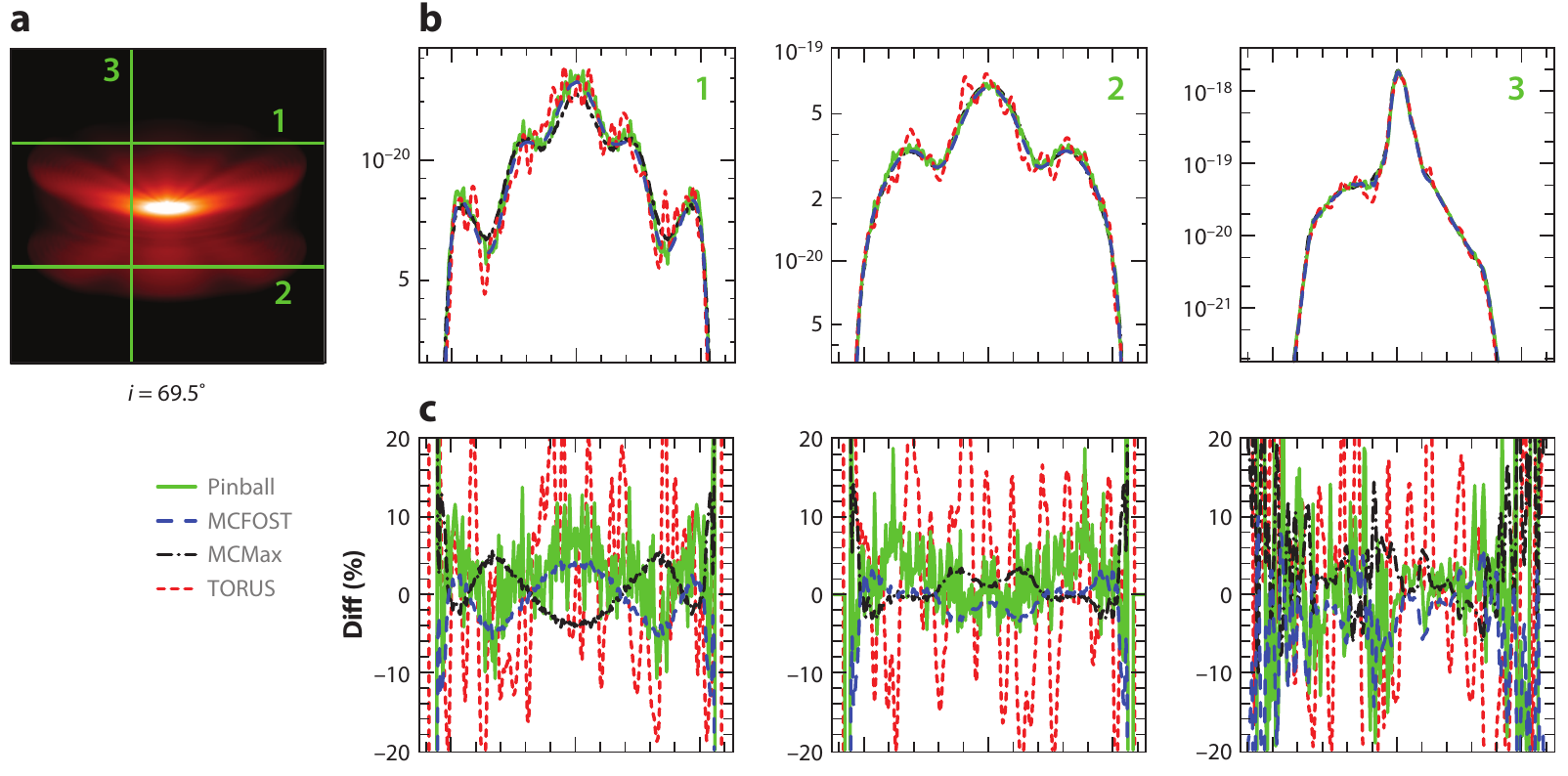}
\caption{An example of the two-dimensional dust radiative transfer benchmark \citep{2009AA...498..967P} for a flared circumstellar disc with a V-band optical depth in the midplane of $\tau = 10^6$. ({\em{a}}) Scattered light image. ({\em{b}}) These three panels show brightness profiles along the cuts plotted in panel {\em{a}}. ({\em{c}}) The differences among the four different models used in the benchmark. The different line styles and colors indicate whether the model was Pinball \citep{2001RMxAA..37..221W}, MCFOST \citep{2006AA...459..797P}, MCMax \citep{2009AA...497..155M}, or TORUS \citep{2004MNRAS.350..565H}.}
\label{PinteBenchmark}
\end{figure}

Probably the most objective way to compare the strengths and weaknesses of the different codes is to use benchmark problems. In the past few years, there have been benchmark efforts in many different computational astrophysics areas, including molecular line transfer \citep{2002AA...395..373V}, halo and void finder algorithms \citep{2008MNRAS.387..933C, 2011MNRAS.415.2293K}, astrophysical hydrodynamics \citep{2007MNRAS.380..963A}, cosmological hydrodynamical simulations \citep{1999ApJ...525..554F, 2005ApJS..160....1O}, and cosmological radiation hydrodynamics \citep{2006MNRAS.371.1057I, 2009MNRAS.400.1283I}. At the moment, no code validation or benchmark project exists for 3D dust RT. The most advanced dust RT code validation projects are 2D benchmarking efforts.

\citet{2004AA...417..793P} presented a benchmark test for 2D equilibrium RT problems. Their system consists of a single star surrounded by a flared axisymmetric accretion disc. The optical depth through the disc varies from $\tau_{\text{V}} = 0.1$ to $\tau_{\text{V}} = 100$. The accretion disc contains strong density gradients, which makes it an ambitious benchmark problem (unfortunately, anisotropic scattering is not taken into account). The authors compare the temperature maps and SEDs for five 2D dust RT codes (two grid-based codes and three MC codes). Differences between the various codes in the temperature maps are smaller than 1\% for the most optically thin model, but some reach up to 15\% in the most optically thick system. For the emerging spectra, the differences range from a few percent for the optically thin models to more than 20\% for the most optically thick models.

A first extension of the \citet{2004AA...417..793P} benchmark test was presented by \citet{2003Ap&SS.286..113P}, with two of the five codes from the original benchmark participating (one MC code and one RayT code, both intrinsically 3D codes). [The \citet{2003Ap&SS.286..113P} paper was actually a follow-up project.] They considered a similar disc as that in the \citet{2004AA...417..793P} benchmark, with an azimuthal ring added as a simplified model for a spiral density distortion. The main advancements were the addition of anisotropic scattering and that images and visibilities, in addition to SEDs and temperature maps, were compared. The difference in flux in the entire image was smaller than 20\%, but the visibility curves showed substantially larger discrepancies in their overall shapes.

\citet{2009AA...498..967P} go another step further in what is the most advanced dust RT benchmark to date.They start from a similar circumstellar disc model, but consider optical depths up to $\tau_{\text{V}} =10^6$, use anisotropic scattering, and compare images and polarization maps. The four different codes that participate in the benchmark are all MC codes. The agreement in the temperature distributions is very good, with differences almost always smaller than 10\%. Differences in the SEDs remain smaller than 15\% for models with $\tau_{\text{V}} =1,000$ and agree within 20\% over almost the entire wavelength range for the most optically thick cases. Pixel-to-pixel differences in high-resolution scattered light images remain limited to 10\%, and the polarization maps do not differ by more than 5$^\circ$ in regions where the polarization can be effectively measured by observations ({\bf{Figure~\ref{PinteBenchmark}}}).

\section{THE FUTURE OF THE FIELD}
\label{summa}

\subsection{Present Status}

3D dust RT is a rich and diverse field, with applications across a broad range of astrophysical topics from dust near stars to entire galaxies. Correctly modeling the effects of dust on the transfer of radiation is critical to studying many astrophysical objects, including the dust itself. Recent years have seen an impressive improvement in observational capabilities across the electromagnetic spectrum, and this has shown that the dust distribution in many regions is strongly 3D. This requires methods to compute the dust RT that can handle 3D structures and return solutions in a reasonable amount of time. The most common 3D dust RT solver is based on MC techniques, with RayT features in its modern accelerated form. A few applications have used pure RayT solvers. Both methods face the challenges of grid discretization, determination of uncertainties in solutions, and accurate comparison between observations and the model calculations. Almost 30 codes are currently able to deal with the full 3D dust RT, with code variations arising from the prime field of application. There is no 3D dust RT benchmark; currently, code comparisons are done using 2D benchmarks.

\subsection{General Trends}

Several trends indicate that the future of 3D dust RT is bright. The number of people actively involved in 3D dust RT is growing, and the number of new published codes has increased significantly in recent years. A 3D approach to modeling complex distributions is becoming common in many fields requiring 3D dust distributions. The continuing increase in available computing power and storage will support this trend, allowing a full transition from 2D to 3D dust RT for all objects showing 3D signatures. A prominent example of this trend is circumstellar disks with (proto)planets, where the MHD simulations have been 3D for years, dust RT modeling often was 2D, and observations are now reaching the resolution necessary to identify the 3D signatures of disk deformation due to a planet. In addition, modern online tools are expected to support the access to the codes by users through sophisticated interfaces.

\subsection{Future Benchmarks}

For progress in 3D dust RT to continue, 3D dust RT benchmarks need to be established. Given the complexity of the codes, increasing number of acceleration algorithms, and large number of specific applications, it is critical to provide a quantitative comparison between codes. Experience with existing dust RT benchmarks and similar efforts in other areas indicate a suite of 3D dust RT benchmarks is needed. Ideally, each benchmark would focus on a particular part of the RT solution (e.g., scattering, polarization, equilibrium dust emission, or nonequilibrium dust emission) in a 3D geometry. This would provide a clear test of different aspects of 3D dust RT and support the participation of all codes in at least part of the suite.

\subsection{Data Modeling Future}

Given the impressive flow of new data from ground- and space-based observatories now and projected for the coming years, it is clear that the demand to accurately model 3D dusty structures will rise strongly. Interfaces that can simulate observations with different telescope properties will become necessary to perform modeling. We expect a rise of 3D dust RT modeling efforts that rely on automated fitting processes rather than by-hand explorations of the model parameter space. Because the number of sources of multiwavelength data will rise, collaborations between observers and modelers will become more frequent. The ultimate goal of 3D dust calculations is to model multiwavelength images and derive quantitative and statistically sound information about 3D structures, embedded sources, and the dust itself.

\subsection{Future Connections to Nondust Radiative Transfer Codes}

Another future direction is the coupling of 3D dust RT codes with codes describing other physical effects in astrophysical objects. This trend is already happening with 2D dust RT codes, and the extension to 3D dust RT codes is clearly the next step. A variant of this type of connection is already happening where 3D dust RT is used to calculate the radiation field in a dust distribution generated with an MHD code; furthermore, MHD codes that make use of simple dust RT could be tested or the simple algorithms improved by comparison with full 3D dust RT solutions. Chemical network calculations could be based on a more realistic estimate of the incoming radiation calculated from 3D dust codes. Finally, a combined calculation for 3D line and dust RT would enable line and continuum data to be simultaneously investigated using the same underlying physical model.

\subsection{Future Algorithms}

Conferences and keyword-related publication searches have often been used in the past to improve the unfortunately rare communication of new numerical algorithms from applied mathematics to astrophysics. The basic issue is the sheer flow of new findings and the different languages of the two communities. Recent MC improvements have been developed mainly by coders working in the field, and additional efforts should be made to enable community-crossing exchange on algorithms and error control. As a result of communications between coders preferring different solvers, we expect hybrid solvers making use of the advantages of the various approaches to appear more frequently. Given the increase in complexity in the modeled objects, we expect future activities to establish grid generation algorithms that are optimized for 3D dust RT; besides the octree or AMR-style grids that are now routinely implemented in 3D dust RT codes, unstructured grids as used in line RT \citep{2010A&A...515A..79P}, and MHD codes \citep{2010MNRAS.401..791S} are an interesting alternative. The inclusion of time dependence in the 3D RT problem, which could be important in the context of star formation or episodic accretion, will also need to be tackled with new algorithms \citep[see, e.g.,][]{2011MNRAS.416.1500H}. The increasing availability of massively parallel machines will support algorithms that are optimized to run on many processors.

\subsection{Input Physics Improvements}

The improvement of the solvers is not restricted to developing algorithms that provide accelerated solutions. The interaction of radiation with cosmic dust is still not fully understood, and the variation of the dust properties with environment is an area of active research. The various continuum radiation sources such as stars, PDRs, AGN accretion discs, and the interstellar radiation field are areas of vigorous investigation. For example, efforts based on existing and upcoming large-scale surveys are being made to update the 3D structure of the stars in the Milky Way. Consequently, we expect to achieve a better understanding of the observed radiation from future research on the optical properties of dust and improved data on the stellar and nonstellar sources that enter the 3D dust RT equation.

\subsection{Challenges}

A major challenge in 3D dust RT that this review highlights is how to account for and mitigate systematic uncertainties in the dust RT solution. They arise from under-resolving grids, not propagating rays/photons to important cells, and/or uncertainties in the underlying dust grain models. As under-resolving of the dust and radiation field grid is often a result of constraints on computer memory and speed, improvements in algorithms to implicitly handle optimal grids are needed. The preprocessing steps necessary for the RayT solver address some of these issues, but need further automating. The issue of not propagating enough rays/photons into particular cells has been solved for both RayT (placement of rays) and MC (biased emission), but both currently require hand-tuning. An algorithm to automatically add additional rays/photons similar to that used for AMR would clearly be useful. Finally, uncertainties in the assumed dust grain model provide a systematic uncertainty in the dust RT modeling that is difficult to quantify. Different dust grain models can be used to provide an estimate of this uncertainty, but the best way to reduce this uncertainty is to support the improvement of dust grain models through the use of improved laboratory and observational data.

\section*{DISCLOSURE STATEMENT}

The authors are not aware of any affiliations, memberships, funding, or financial holdings that might be perceived as affecting the objectivity of this review.

\section*{ACKNOWLEDGEMENTS}

The authors acknowledge the support of the Ghent University for two excellent week-long meetings in Ghent, Belgium, where a large portion of the work on this review was done. We thank Simon Bruderer, Jacopo Fritz, Gianfranco Gentile, Michiel Min, Kirill Tchernyshyov, Ewine van Dishoeck, and Adolf Witt for providing comments on this review that significantly improved it.

\bibliography{RTReview}

\end{document}